\documentclass{aa}  

\usepackage{textcomp}
\usepackage{graphicx}
\usepackage{amsmath}
\usepackage{amssymb}
\usepackage{bm}
\usepackage{upgreek}
\usepackage{IEEEtrantools}
\usepackage{multirow}
\usepackage[dvipsnames]{xcolor}
\usepackage[colorlinks=true,allcolors=blue,urlcolor=blue]{hyperref}
\usepackage{txfonts}
\usepackage[normalem]{ulem} 
\usepackage{booktabs}
\usepackage{longtable}
\usepackage{natbib}
\usepackage{subcaption}
\usepackage{aas_macros}

\newcommand{\sect}[1]{\text{Section~\ref{#1}}}
\newcommand{\fig}[1]{\text{Figure~\ref{#1}}}
\newcommand{\tab}[1]{\text{Table~\ref{#1}}}

\newcommand{\mtd}{\textlangle3D\textrangle}
\newcommand{\mtdmath}{\langle3D\rangle}
\newcommand{\balder}{\texttt{Balder}}
\newcommand{\multitd}{\texttt{Multi3D}}
\newcommand{\scate}{\texttt{Scate}}
\newcommand{\blue}{\texttt{Blue}}
\newcommand{\marcs}{\texttt{MARCS}}
\newcommand{\stagger}{\texttt{STAGGER}}
\newcommand{\atmo}{\texttt{Atmo}}
\newcommand{\graspg}{\texttt{GRASPG}}
\newcommand{\grasp}{\texttt{GRASP}}

\newcommand{\teff}{T_{\mathrm{eff}}}

\newcommand{\lgeps}[1]{\log \varepsilon_{\mathrm{#1}}}

\newcommand{\vmic}{\xi_{\text{mic;1D}}}
\newcommand{\nm}{\mathrm{nm}}
\newcommand{\dex}{\mathrm{dex}}
\newcommand{\kms}{\mathrm{km\,s^{-1}}}

\newcommand{\kelvin}{\mathrm{K}}

\begin{document}

\title{\ion{Ag}{I} model atom and the 3D non-LTE solar silver abundance}

\author{S.~Caliskan\inst{\ref{uu1}},
A.~M.~Amarsi\inst{\ref{uu1}},
P. J\"onsson\inst{\ref{Malmo}},
N.~Grevesse\inst{\ref{liege1},\ref{liege2}},
\and B.~K. Sahoo\inst{\ref{inst}}
}

\authorrunning{Caliskan et al.}

\institute{\label{uu1}Theoretical Astrophysics, 
Department of Physics and Astronomy,
Uppsala University, Box 516, SE-751 20 Uppsala, Sweden\\
\email{sema.caliskan@physics.uu.se}
\and
\label{Malmo}Department of Materials Science and Applied Mathematics, Malm\"o University, SE-205 06, Malm\"o, Sweden
\and
\label{liege1}Centre Spatial de Li\`ege, Universit\'e de Li\`ege,
avenue Pr\'e Aily,
B-4031 Angleur-Li\`ege, Belgium
\and
\label{liege2}Space sciences, Technologies and Astrophysics Research (STAR)
Institute,
Universit\'e de Li\`ege, All\'ee du 6 ao\^ut, 17, B5C,
B-4000 Li\`ege, Belgium
\and
\label{inst}Atomic, Molecular and Optical Physics Division, Physical Research Laboratory, Navrangpura, Ahmedabad 380009, India
}

\date{Received ****; accepted ****}

\abstract{Silver is an important light neutron-capture element whose stellar
abundances can help constrain the origin of the weak r-process. The Sun is an
important reference point for such studies; moreover, being a moderately
volatile element in CI chondrites, the solar silver abundance is interesting as
a diagnostic for the debated Sun–CI abundance versus condensation temperature
trend. These studies require accurate silver abundances, that go beyond the
commonly used assumptions of one-dimensional (1D) atmospheres and local
thermodynamic equilibrium (LTE); however, no consistent 3D non-LTE analysis of
silver has been available to date. We present a new \ion{Ag}{I} model atom built
from carefully curated radiative and collisional data, including newly computed
oscillator strengths using an ab initio multi-configurational Hartree–Fock
method and inelastic hydrogen collision rates based on a combined asymptotic and
free-electron model approach. We assess modelling uncertainties via targeted
sensitivity tests, finding the results to be most sensitive to the hydrogen
collision data. Applying the model to the solar \ion{Ag}{I} $328\,\nm$ and
$338\,\nm$ resonance lines, we find severe positive abundance corrections from
coupled 3D and non-LTE effects. Using revised equivalent width measurements, we
derive a recommended solar 3D non-LTE silver abundance 
of $\lgeps{Ag} = 1.15 \pm
0.08$. This is an increase of $0.19\,\dex$
relative to the current reference
value.  Our ab initio model significantly 
reduces the discrepancy
with the meteoritic value from $0.25\,\dex$ to
$0.06\,\dex$; moreover, this residual offset
is consistent with what has recently been reported for
other moderately volatile elements.
The Sun provides
the benchmark test for the first silver non-LTE model atom presented here. In
subsequent work, this model will be applied to determine 3D non-LTE silver
abundances in metal-poor dwarfs and giants, enabling improved constraints on
Galactic chemical evolution and weak r-process nucleosynthesis.}


   \maketitle
%

\section{Introduction}

Silver is a light neutron-capture element (Z=47) whose astrophysical origin, and
the relative yields of its different production processes and sites, have long
been of interest. Yet stellar constraints on silver remain sparse, as only two
\ion{Ag}{I} resonance lines are observable, and both lie in the near-UV where
they are strongly affected by blends.
\cite{hansen_origin_2011} and \cite{hansen_silver_2012} are recognised as the first publications where the silver abundance was analysed for a large sample of stars. Even before these papers, several studies \citep[e.g.][]{2006ApJ...643.1180H, francois_first_2007, sneden_neutron-capture_2008, 2010ApJ...724..975R} found that lighter r-process elements (i.e. 34 $\le$ Z $<$ 56) exhibit a departure from the predicted solar system r-process residual pattern, compared to the heavier r-process elements exhibiting a universal r-process distribution. This suggests that more than one r-process channel contributes, commonly referred to as a ``weak' and a main channel \citep[e.g.][]{wanajo_r-process_2006,ott_meteoritic_2008,kratz_explorations_2007,montes_nucleosynthesis_2007, farouqi_nucleosynthesis_2009}.

Silver can be considered a tracer of the weak r-process. First, it is produced predominantly by the r-process: 
\citet{1999A&A...342..881G}, \citet{sneden_neutron-capture_2008}, and \citet{prantzos_chemical_2020} predict that nearly 80$\%$ of solar-system silver originates from r-process nucleosynthesis. Moreover, silver lies near the middle of the atomic-number range where the lighter r-process elements deviate from the main solar-scaled r-process distribution, making it well placed to probe any ``weak'' component. In their paper, \cite{hansen_silver_2012} supported this interpretation by showing a strong anti-correlation at low metallicities between silver and europium (formed by the main r-process). Their findings were later confirmed by \cite{wu_palladium_2015}, who increased the sample of stars with observed silver abundances to solar and super-solar metallicities, and more recently by \cite{huang_palladium_2025}, who extended it to extremely metal-poor stars. However, these studies all report large star-to-star scatter in the silver trends, and \cite{hansen_origin_2011} additionally found an abundance difference of around $0.5\,\dex$ between dwarfs and giants at similar metallicity.  

Silver in our solar system also exhibits a puzzling discrepancy. The currently adopted solar photospheric value is that of \cite{grevesse_elemental_2015},  
$\lgeps{Ag}=0.96\pm0.10$\footnote{$\lgeps{Ag}\equiv\log_{10}N_{\mathrm{Ag}}/N_{\mathrm{H}}+12$.},
which was derived via synthesising the two \ion{Ag}{I} UV resonance lines using a three-dimensional (3D) radiation-hydrodynamical model atmosphere.  However, as discussed in \cite{2021A&A...653A.141A}, the present-day solar photospheric abundance of silver differs greatly from that inferred from CI chondrites, with the value of \citet{grevesse_elemental_2015} being $0.25\,\dex$ lower than the meteoritic value from \cite{lodders_meteoritic_2021}.

All of the aforementioned studies have derived silver abundances under the assumption of local thermodynamic equilibrium (LTE), be it in 1D for large samples of stars \citep[e.g.][]{hansen_origin_2011,hansen_silver_2012,wu_palladium_2015,huang_palladium_2025}, or in 3D for the Sun \citep{grevesse_elemental_2015}. Thus, the potential impact of 3D non-LTE effects on the star-to-star scatter and the dwarf–giant offset in metal-poor stars, as well as on the inconsistency between the solar photospheric silver abundance and that found in primitive meteorites, has not yet been assessed.   Detailed 3D non-LTE modelling is therefore needed to clarify the formation channels of silver and its role as a weak r-process tracer, as well as the true solar reference abundance.

To this day, there are no non-LTE calculations of silver in the literature, likely in part because of incomplete atomic data for silver. Non-LTE modelling requires a so-called model atom, containing extensive radiative and collisional data. In this paper, we compile the best available atomic data for silver that are needed for non-LTE spectrum synthesis, and compute missing ingredients where necessary, such as for the inelastic collisions with neutral hydrogen (Section~\ref{method}).  We then investigate the 3D and non-LTE effects on the formation of the \ion{Ag}{I} $328\,\nm$ and $338\,\nm$ resonance lines in the solar photosphere (Section~\ref{results}).  We calculate the 3D non-LTE abundance correction for the Sun, and re-evaluate the solar silver abundance in comparison with previous determinations (Section~\ref{discussion}). Our goal is to benchmark the model atom on the Sun, prior to calculating and applying 3D non-LTE abundance corrections to other late-type stars, which will be addressed in future work (Section~\ref{conclusion}).

\section{Methods}
\label{method}

\subsection{Model atom}\label{model_atom}

\begin{figure}
\centering
\includegraphics[width=\hsize]{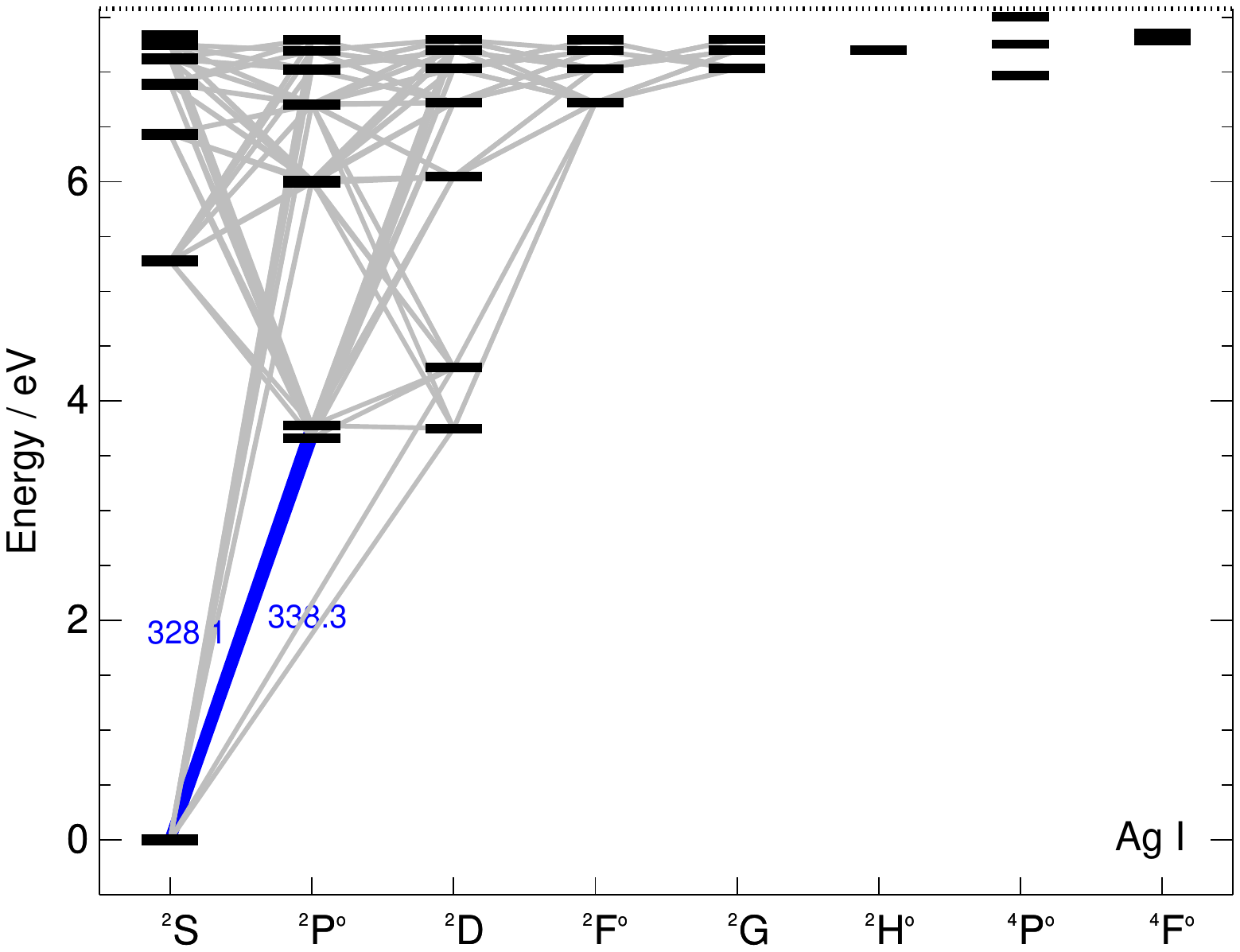}
\caption{Grotrian diagrams for \ion{Ag}{I} illustrating the model atom. The transitions highlighted in blue correspond to the two \ion{Ag}{I} diagnostic lines analysed in this work (vacuum wavelengths). The dotted horizontal line marks the silver ionization limit.
\label{fig:model}}
\end{figure}

A model atom for the non-LTE modelling of \ion{Ag}{I} was constructed for this study. The structure of the model is illustrated in \fig{fig:model}. We present an overview of the different ingredients used in the model here.

The model consists of 57 energy levels with fine structure, taken from the NIST database \citep{NIST_ASD}.  This includes five fine-structure levels of \ion{Ag}{II}, namely the ground state, as well as the first two excited terms. For the bound-bound radiative transitions, only seven oscillator strengths ($f$ values) can be found on NIST.

This set of seven oscillator strengths from NIST were then complemented with values for 53 more transitions via ab initio combined multiconfigurational Dirac–Hartree–Fock and relativistic configuration interaction (MCDHF+RCI) calculations performed using the \graspg{} package \citep{SI2025109604}, an extension of \grasp{} \citep{FROESEFISCHER2019184}, for the important low-lying levels up to $4d^{10} 8s \, ^2\mathrm{S_{1/2}}$ ($6.89\,\mathrm{eV}$). Further details about the methods and results of this calculation may be found in \citet{2026A&A...709A..31J}. In the paper, the computed excitation energies and transition rates were compared to results from Fock-space multi-reference coupled cluster (FSMRCC) calculations \citep[e.g.][]{sahoo_coupledcluster_2025}, and generally a good agreement was found. Moreover, the MCDHF+RCI lifetimes were in agreement with experimental lifetimes. 

For the two diagnostic lines, we adopted the oscillator strengths from NIST, with $\log gf = - 0.046$ and $\log gf = - 0.356$ for the 328 nm and 338 nm line, respectively. These values agree with the experimental values of $\log gf =-0.021$ and $-0.333$ from \cite{5944TP}. They also are in close agreement with the calculated MCDHF+RCI ($\log gf = - 0.036$ and $\log gf = - 0.352$) and FSMRCC ($\log gf = - 0.003$ and $\log gf = - 0.315$) values.

For the transitions including higher levels, oscillator strengths were taken from \cite{civis_time-resolved_2010}, calculated using the Fues model potential approach (FMP). A comparison between the two sets of oscillator strengths and their impact on the non-LTE abundance can be found in Section~\ref{discussion_sensitivity}.

Natural broadening coefficients were calculated via: $$\gamma_{ul} = \sum_{l'<u} A_{ul'} + \sum_{l'<l} A_{ll'},$$ where $A_{ul'}$ and $A_{ll'}$ are transition rates from the upper and lower level of the $ul$ transition, respectively, to all other lower levels \citep[e.g.][]{gray_book_2022}.
Pressure broadening due to hydrogen collisions were obtained by interpolating the tables of Anstee, Barklem, and O’Mara (ABO theory; \citealt{barklem_2000A&AS..142..467B}) when available, and the classical Uns\"{o}ld theory was applied otherwise, adopting a fudge factor of $2.0$ (see the discussion in \citealt{barklem_review_2016}). In addition, hyperfine structure and isotopic contributions were added for the two diagnostic lines. The same values were adopted as in \cite{hansen_silver_2012}, where they derived new hyperfine components, instead of using those from \citet{1972SoPh...25...30R}, as commonly used in abundance studies of the \ion{Ag}{I} resonance lines in the literature \citep[e.g.][]{wu_palladium_2015,huang_palladium_2025}, since they only included two hyperfine components per isotope, missing one component.
The adopted isotopic ratio is $51.84\%$ for isotope $107$ and $48.16\%$ for isotope $109$ \citep{2021A&A...653A.141A}. 
Nevertheless, we found that the hyperfine and isotopic splitting affects the non-LTE equivalent widths of the two diagnostic lines by only around $0.01\,\dex$ in the solar photosphere.

Photoionisation cross-sections were estimated using the hydrogenic approximation, using the Kramers formula \citep[e.g.][]{rutten_radiativetransfer_2003}: $$\sigma = 2.815 \times10^{29}\frac{Z^4}{n_{\mathrm{eff}}\nu^3},$$
in units of $\mathrm{cm}^{2}$, where $n_{\mathrm{eff}}$ is the effective principal quantum number and $\nu$ is the frequency. The Gaunt factor was set to unity.

The rate coefficients for excitation via electron collisions were calculated using the recipe of \cite{seaton_impact_1962} for the transitions with available transition rates $A$. For the other lines, the semi-empirical van Regemorter recipe \citep{van_regemorter_rate_1962} was used, by assuming an effective oscillator strength of $0.001/g_{l}$, with $g_l$ the statistical weight of the lower level, to estimate the Einstein coefficients for spontaneous emission entering the recipe. For ionisation via electron collisions, the empirical formula given in \citet{allen_astrophysicalquantities_1973} was adopted.

Finally, for excitation and charge transfer via neutral hydrogen collisions, we performed calculations based on the asymptotic model approach described in \cite{Barklem_2016PhRvA..93d2705B, Barklem_2017PhRvA..95f9906B} for the rate coefficients, for all the energy levels included in the model atom (except h states). For the parentage coefficients, appearing in the expression of the wavefunction for the Ag + H quasi-molecule used to calculate the potentials and couplings, we adopted the same values calculated for \ion{Cu}{I} in \cite{2025A&A...696A.210C} since they are homologous elements and therefore have very similar energy structures. We combined the asymptotic model rates with excitation rate coefficients calculated via the code from \cite{Barklem_2017ascl.soft01005B}, which is based on the free electron model by \cite{Kaulakys_1991JPhB...24L.127K}, for all \ion{Ag}{I} energy levels. It was shown in the literature that the latter approach may better account for the transition mechanisms involving highly excited levels near the relevant ionic limit and for states where the avoided crossing occurs at long and short internuclear distances \citep[e.g.][]{2018A&A...616A..89A}. Combining the two methods therefore prevents underestimating the hydrogen collisions. Since these two approaches do not account for fine-structure, we redistributed the rate coefficients among fine-structure levels by dividing the coefficients calculated without fine-structure by the total number of final states, following Boltzmann distributions. Finally, we also assign very large rate coefficients for transitions between fine-structure levels, motivated by the Massey criterion \citep{Massey_1949RPPh...12..248M}, which suggests that inelastic collisions efficiently couple these levels referred to as ``relative LTE''.

As can be seen from the model atom in \fig{fig:model}, the energy levels belonging to the $4d^9 5s 5p$ configuration in the quartet spin system are included in the model atom, but they are not connected by bound-bound radiative transitions in the model. The transitions connecting the quartet system to the $4d^{10}$ levels in the doublet system correspond to two-electron transitions and thus are expected to be intrinsically weak, as well as the spin-forbidden transitions connecting the quartet system to the $4d^9 5s^2$ levels in the doublet system. In fact, to our knowledge, there are no oscillator strengths for these transitions available in the literature, likely because the lines are too weak to measure, and because the atomic structure is challenging to calculate accurately (these levels are known to perturb the energy structure, as discussed for the homologous system Au I in \citealt{2024JPhB...57e5003C}).  Consequently, in the present model, these transitions couple to the doublet system only via collisional transitions, as well as to the excited levels of \ion{Ag}{II} via both collisions and photoionisations. Furthermore, the two $^2\mathrm{H}$ levels are connected to the rest of the levels only via collisions/photoionisations as well, due to a lack of oscillator strengths for these transitions.

\subsection{Model atmospheres}

\begin{figure}[htbp]
  \centering
  \includegraphics[width=0.87\linewidth]{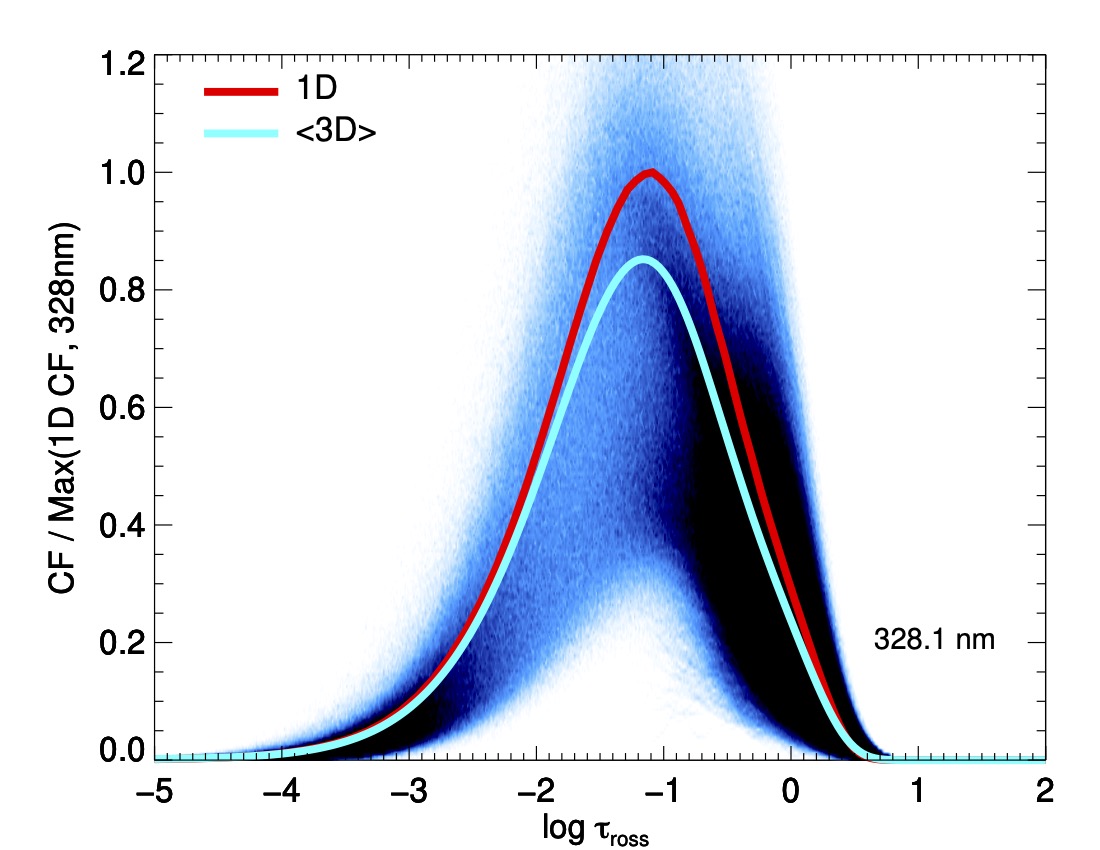}\par\medskip
  \includegraphics[width=0.87\linewidth]{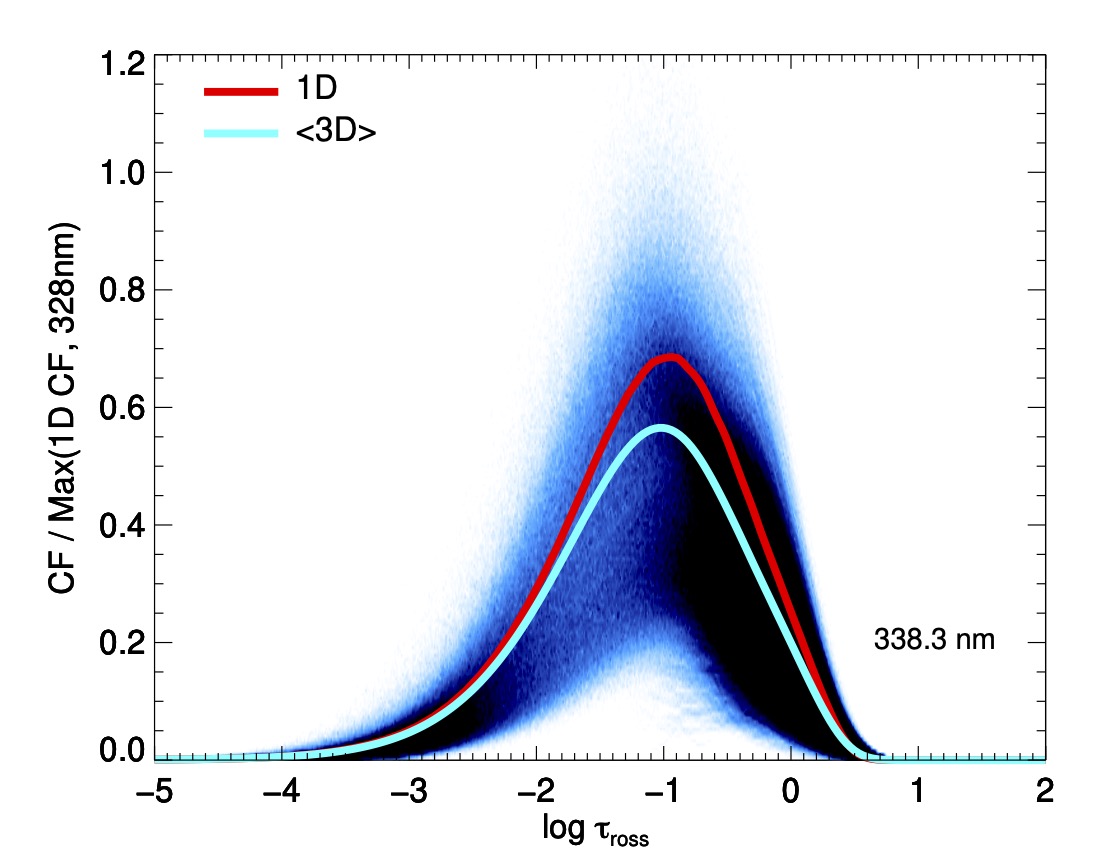}\par\medskip
  \includegraphics[width=0.87\linewidth]{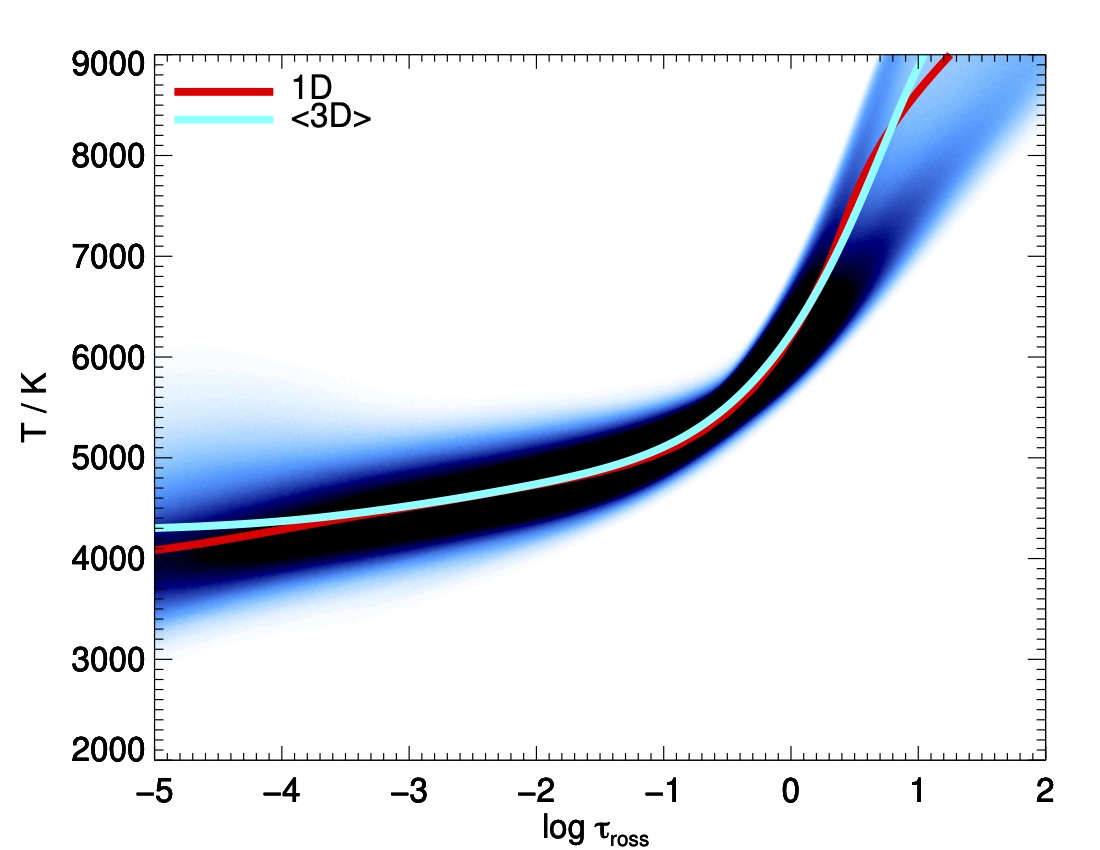}
  \caption{Top and middle panels: non-LTE contribution functions (CF) to the line depression in the vertical intensity \citep[e.g.][]{2015MNRAS.452.1612A} for the two diagnostic \ion{Ag}{I} lines, as a function of vertical optical depth. The contours show the distributions in the 3D model solar atmosphere, and the CF in 1D and \mtd{} are overplotted. All are normalised to the maximum of the 1D LTE CF of the $328\,\nm$ line.
  Bottom panel: the temperature stratification of the 1D, \mtd{} and 3D model solar atmospheres.}
  \label{fig:atmos}
\end{figure}

We illustrate the different model solar atmospheres used in this  work in the lower panel of \fig{fig:atmos}. The 3D model atmosphere is the same one as was first described in \citet{2018A&A...616A..89A}.  In brief, it is a 3D radiation-hydrodynamics simulation that was calculated using the \stagger{} code \citep[e.g][]{2018MNRAS.475.3369C,2024ApJ...970...24S}.  The model is wide enough to encompass around $10$ granules at any given time, and spans around six pressure scale heights below the optical surface and eight pressure scale heights above it, with a resolution that is comparable to that of the \stagger{}-grid \citep{2013A&A...557A..26M}.  The parameters of the model are close to the standard solar values \citep{prsa_sun_2016}, with the gravitational acceleration corresponding to $\log g=4.44\,[\mathrm{cm\,s^{-2}}]$ and a time-averaged effective temperature of $\teff=5773\,\kelvin$ with a snapshot-to-snapshot standard deviation of $16\,\mathrm{K}$.  The model was computed using the solar elemental abundances of \citet{asplund_sun_2009}.  In the subsequent post-processing calculations (Section~\ref{method_balder}), the silver abundance was allowed to vary freely, under the assumption that it is a trace element with no influence on the background pseudostatic model atmosphere.

In order to obtain differential abundance corrections, we also performed 1D calculations using a 1D version of the 3D \stagger{} model, calculated using the \atmo{} code (see Appendix A of \citealt{2013A&A...557A..26M}).  This was constructed using the same equation of state, opacity binning, and solar parameters as for the 3D model.

Finally, we also performed calculations on the temporally- and  horizontally-average of the 3D model solar atmosphere. Details of its construction can be found in \cite{2018A&A...616A..89A}. We refer to this as the mean 3D ($\langle \rm{3D} \rangle$) model hereafter. This is useful for decomposing the 3D effect into a granulation effect and a stratification effect \citep[e.g.][]{lind_review_2024}, as it will be discussed in Section~\ref{3d}. Moreover, the sensitivity tests in Section~\ref{discussion_sensitivity} are based on this model rather than on the full 3D model, to save computational time.

\subsection{Line formation calculations and abundance corrections}
\label{method_balder}

The statistical equilibrium as well as the synthetic emergent spectra were calculated using the 3D non-LTE radiative transfer code \balder{} \citep{2018A&A...615A.139A}, which is a custom version of \multitd{} \citep{1999ASSL..240..379B, Leenaarts_2009ASPC..415...87L}.  Using the code \blue{} \citep{2016MNRAS.463.1518A,2023A&A...677A..98Z},
the equation-of-state and background continuous opacities were calculated on the fly, whereas background line opacities were pre-computed and interpolated onto the model atmosphere at run-time. The non-LTE iterations used $26$ rays on the unit sphere based on the Lobatto quadrature for the polar angle and a trapezoidal quadrature for the azimuthal angle \citep{2024A&A...690A.128A}.  While this study is primarily based on disc-centre intensities, disc-integrated fluxes were also calculated, integrating over $41$ rays on the unit hemisphere.

For the 3D model atmosphere, calculations were performed on 10 snapshots. As in previous work  \citep[e.g.][]{2018A&A...616A..89A}, these were downsampled by a factor of three in each of the horizontal dimensions and refined in the vertical dimension, prior to post-processing with \balder{}.  For the 1D and \mtd{} models, microturbulent broadening was introduced and set to $1.0\,\kms$.  No microturbulent broadening was used for the 3D calculations, as these effects primarily reflect the granulation contrast rather than true turbulent broadening  \citep[e.g.][]{1997A&A...328..229N} and so are naturally included in the 3D radiative transfer calculations with \balder{}.

Emergent spectra were calculated for seven different abundances of silver, from $-0.6\,\dex$ to $+0.6\,\dex$ in steps of $0.2\,\dex$, without the inclusion of background line opacities.  This was also repeated without foreground (silver) line opacities, so as to determine the theoretical continuum and thereby arrive at continuum-normalised spectra.  Equivalent widths were determined by direct integration across the normalised, unblended \ion{Ag}{I} $328\,\nm$ and $338\,\nm$ lines.

We use the notation 
$\mathrm{\Delta^{I}(3N-1L)}$ to indicate the 3D non-LTE ($\mathrm{3N}$) versus 1D LTE ($\mathrm{1L}$) abundance correction for the disc-centre intensity ($\mathrm{I}$).
The abundances were determined via spline interpolation of the silver abundance as a function of the calculated logarithmic equivalent width, onto the input logarithmic equivalent width measured at the solar disk centre (listed in \tab{tab:abundances}). We use similar notation for other abundance corrections, for example $
\mathrm{\Delta^{I}(3N-3L)}$ for the 3D non-LTE versus 3D LTE abundance correction.

\section{Results}
\label{results}

\begin{table*}
\centering
\caption{\ion{Ag}{I} abundances inferred from different spectrum synthesis models.}
\label{tab:abundances}

\begin{tabular}{@{}lccccccc@{}}
\toprule
\toprule
\multirow{2}{*}{Line wvl (nm)} & \multicolumn{7}{c}{$\lgeps{Ag}$} \\
\cmidrule(lr){2-8}
    & W (pm) & 3D non-LTE & 3D LTE & \mtd{} non-LTE & \mtd{} LTE & 1D non-LTE & 1D LTE \\
\midrule
    \ion{Ag}{I} 328.1 & $3.06\pm 0.36$ & 1.173 & 0.892 &
    1.143 & 0.935 & 
    1.086 & 0.872 \\
    \ion{Ag}{I} 338.3 & $2.18\pm 0.17$ & 1.151 & 0.882 &
1.143 & 0.946 & 
    1.089 & 0.886 \\

\midrule
    Arithmetic mean $\lgeps{Ag}$ & & 1.162 & 0.887 & 
    1.143 & 0.941 & 
    1.087 & 0.879  \\
    Recommended $\lgeps{Ag}$ & & $1.15 \pm 0.08$  & & & & &  \\

\bottomrule
\end{tabular}
    \tablefoot{Based on an analysis of the disc-centre intensity. The uncertainty on the recommended abundance is based on measurement and modelling uncertainties added in quadrature
    (see Section~\ref{sun}).}
\end{table*}

\subsection{Non-LTE effects}\label{nlte}

\begin{figure*}[htbp]
    \centering
    \includegraphics[width=0.24\textwidth]{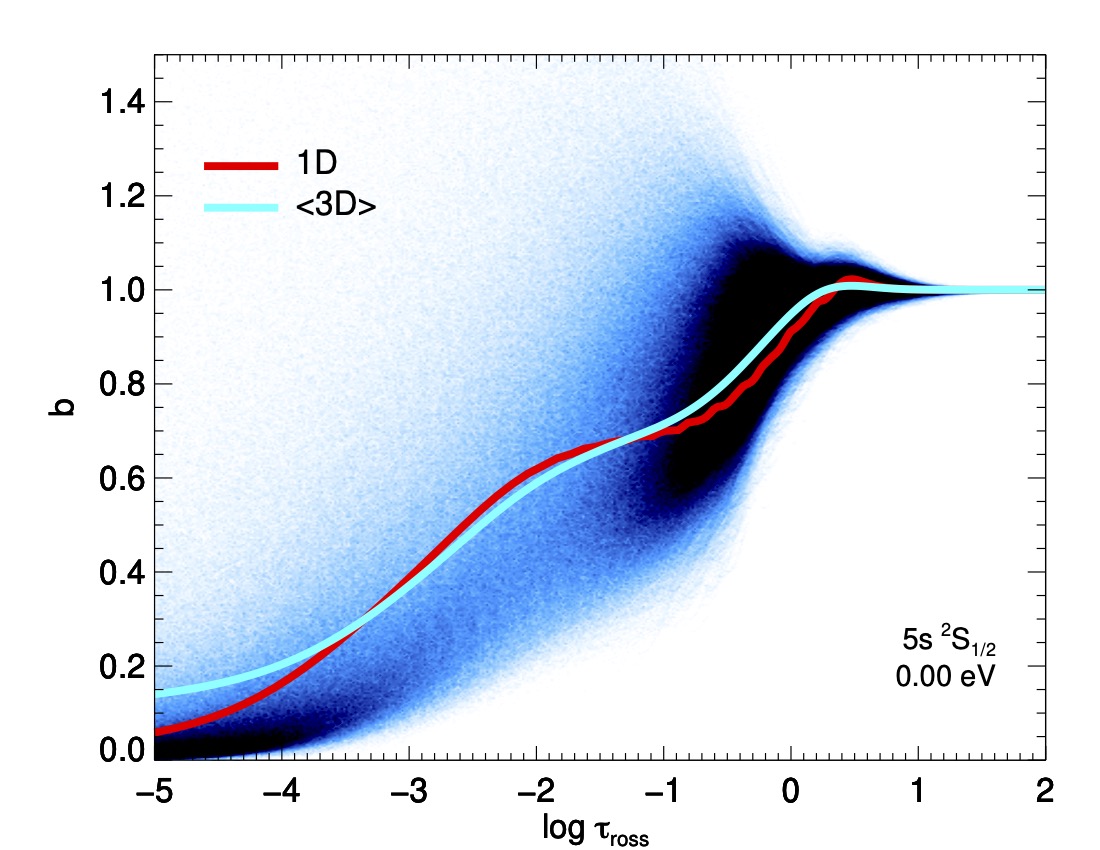}
    \includegraphics[width=0.24\textwidth]{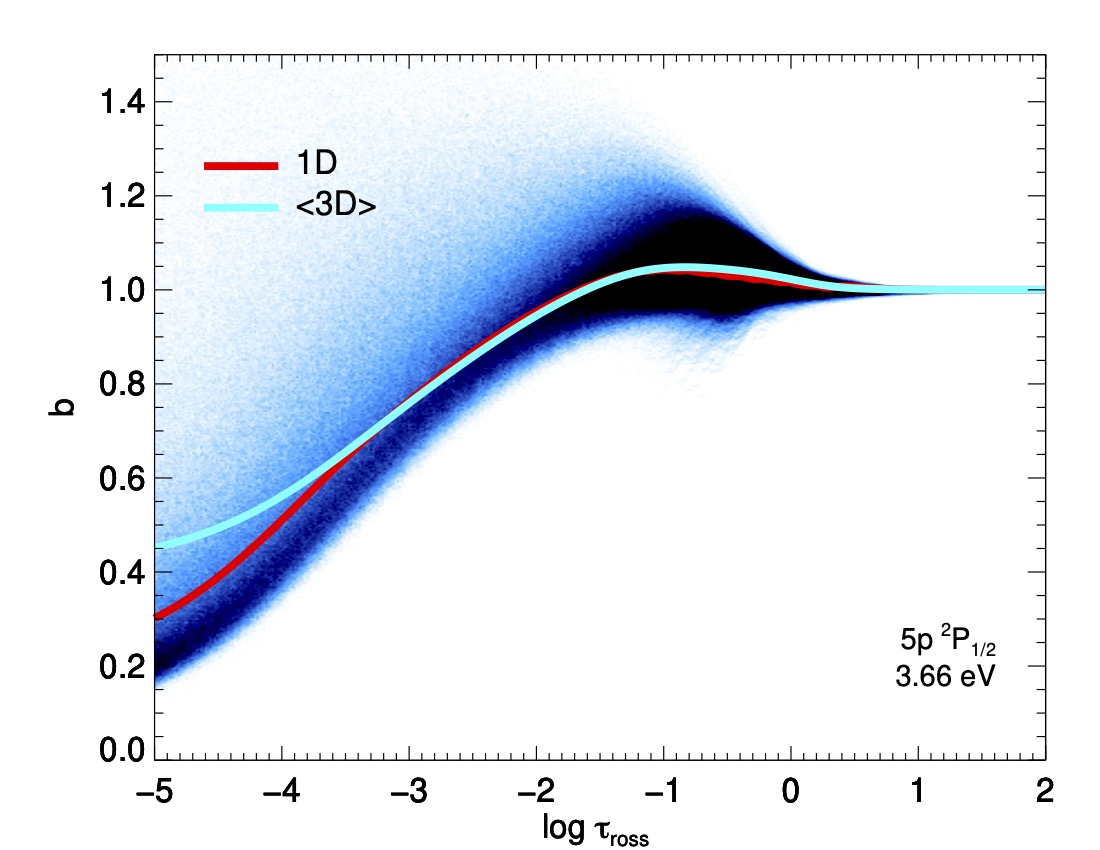}
    \includegraphics[width=0.24\textwidth]{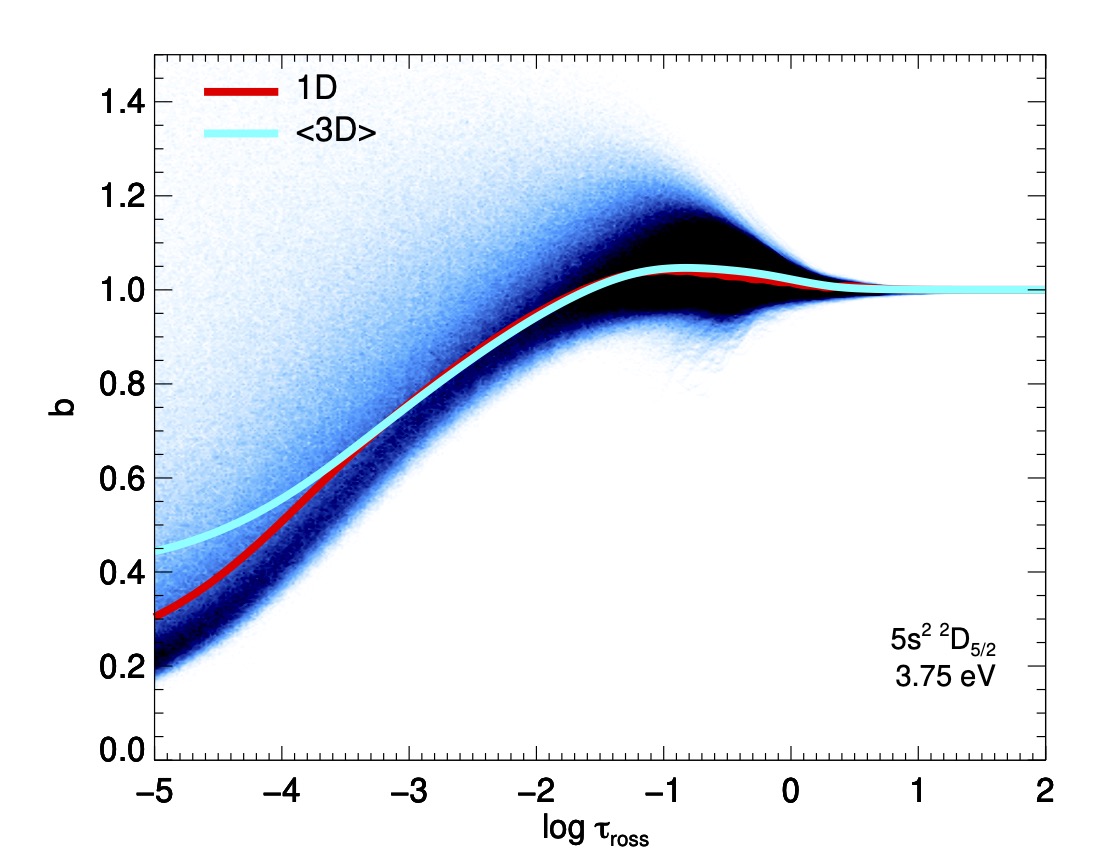}
    \includegraphics[width=0.24\textwidth]{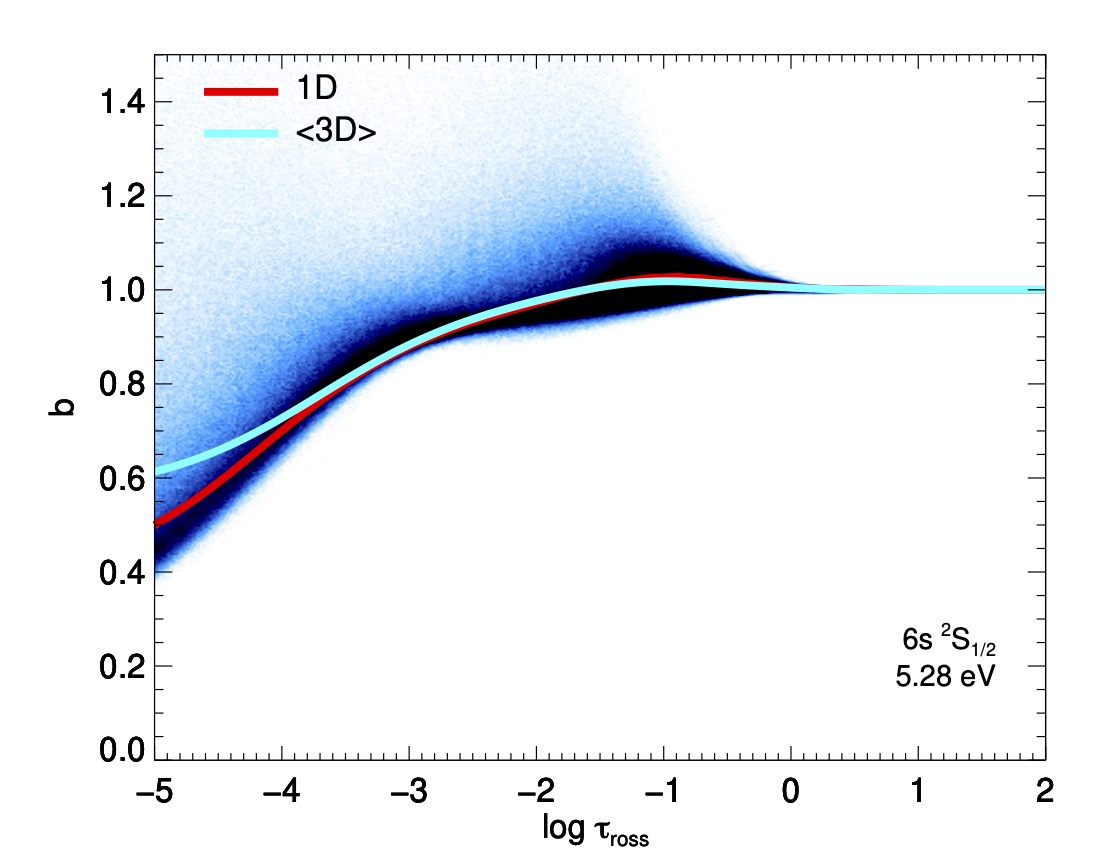}\\[2mm]
    \includegraphics[width=0.24\textwidth]{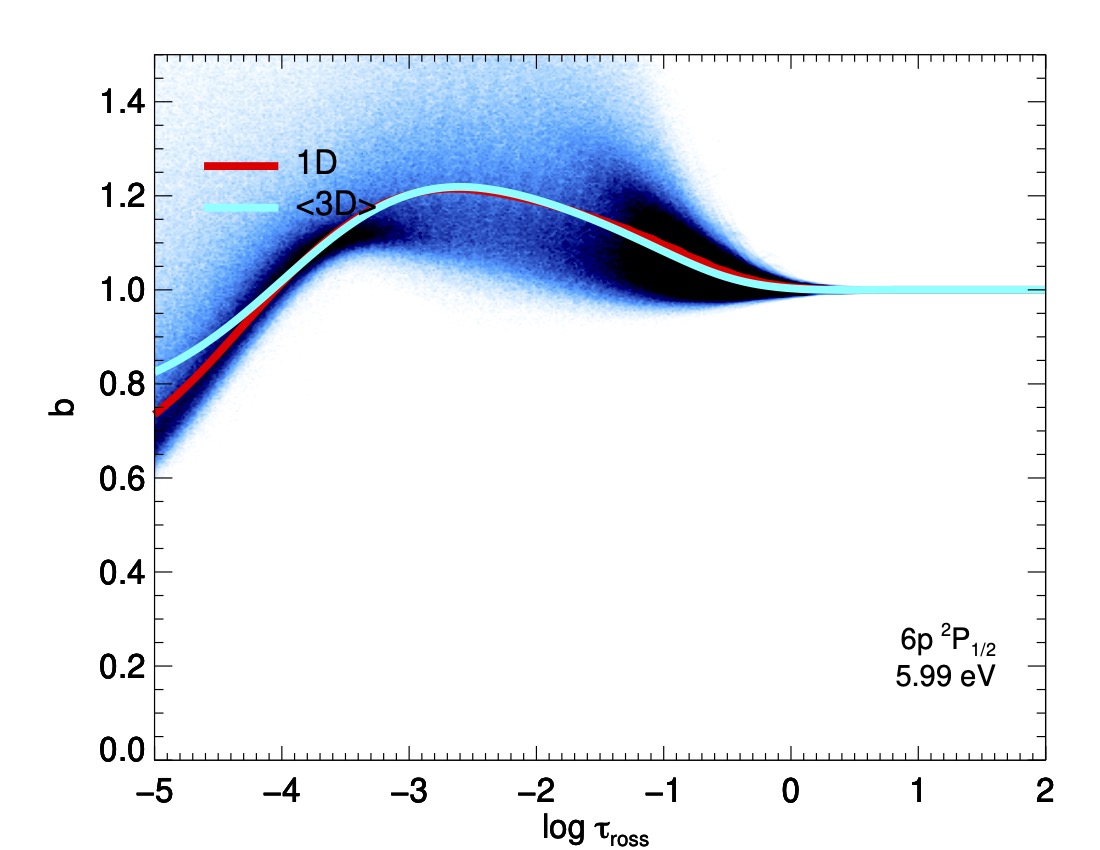}
    \includegraphics[width=0.24\textwidth]{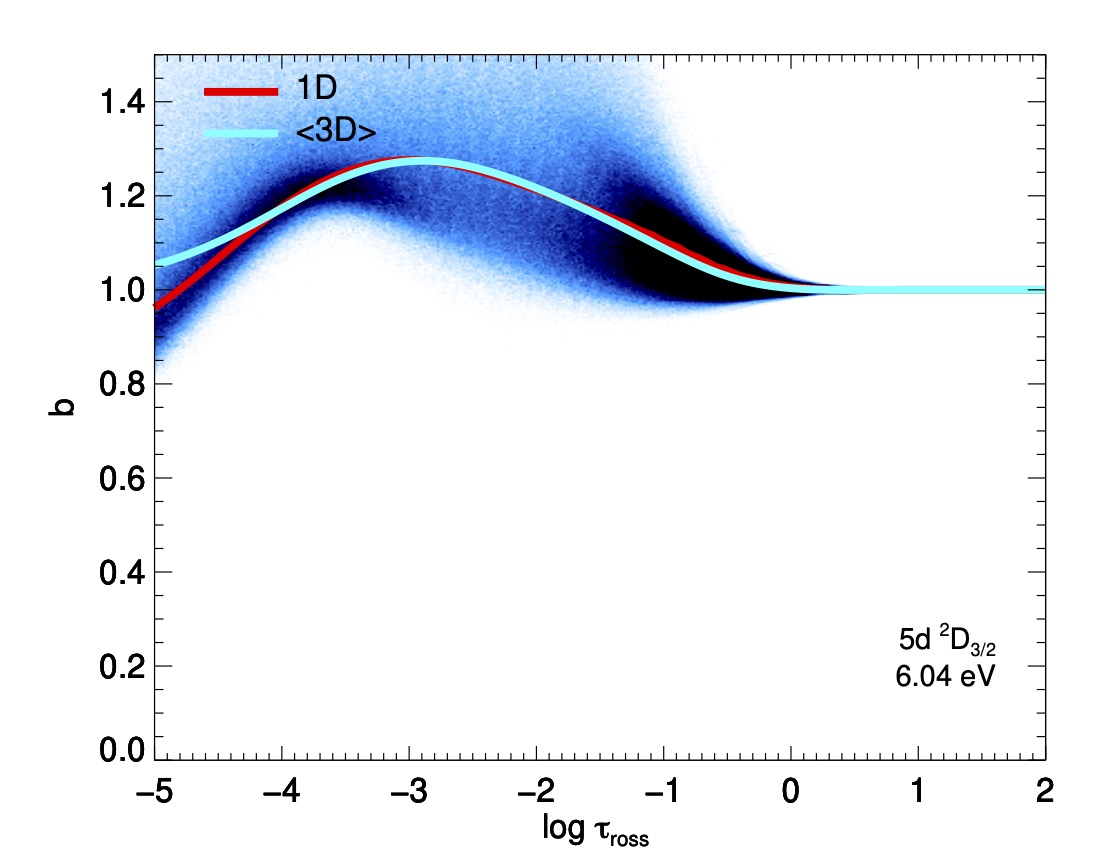}
    \includegraphics[width=0.24\textwidth]{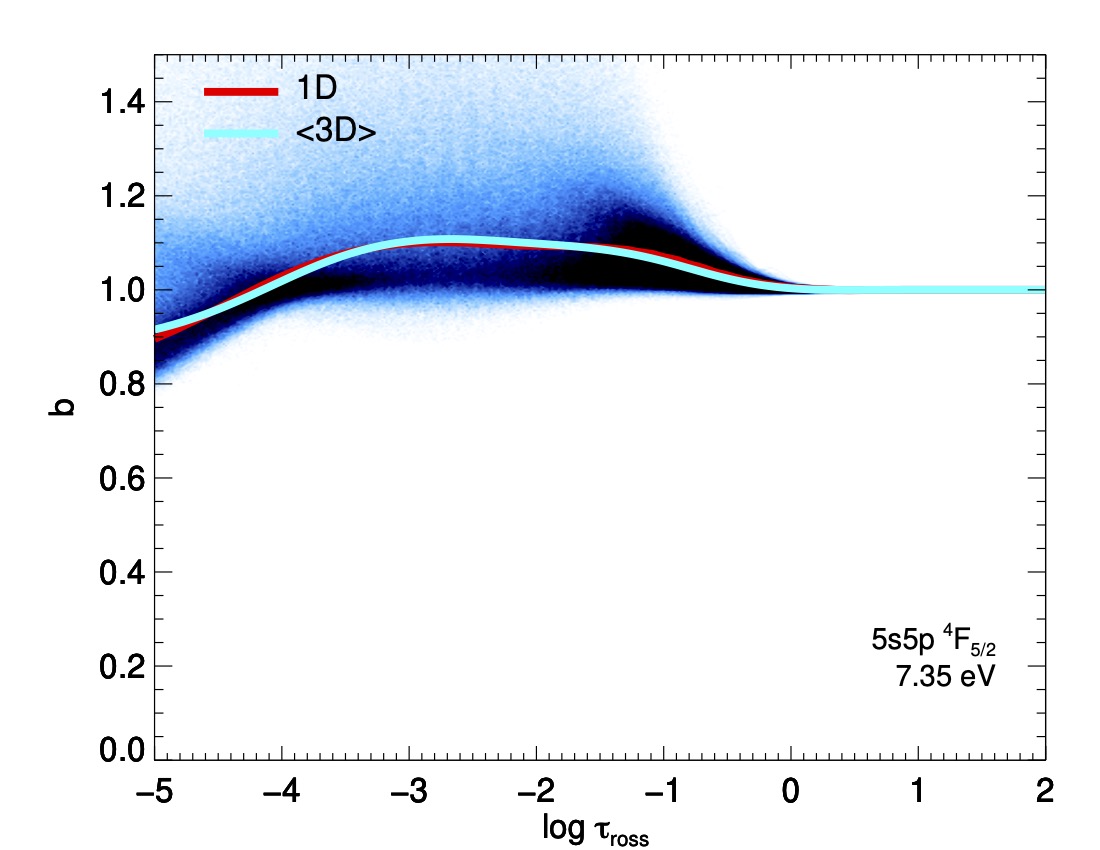}
    \includegraphics[width=0.24\textwidth]{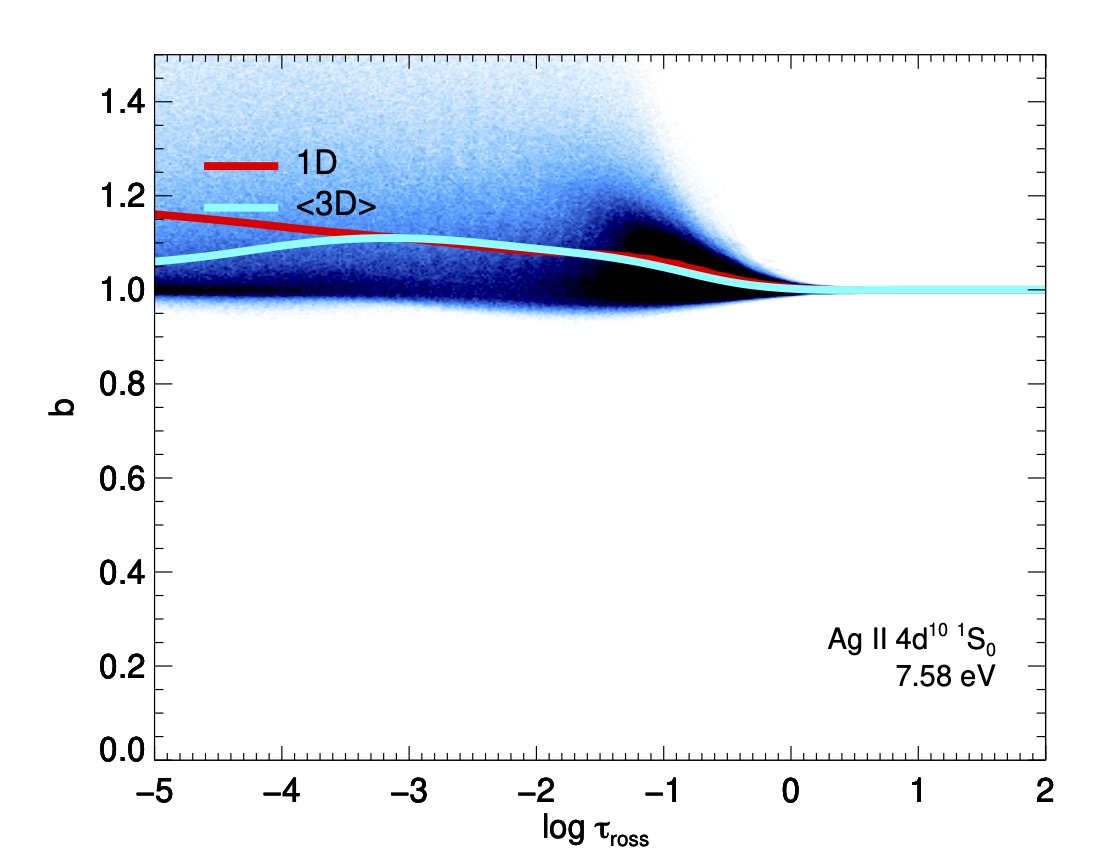}
    \caption{Departure coefficients for seven energy levels of \ion{Ag}{I} (in increasing excitation energy from left to right) and the ground level of \ion{Ag}{II}. The contours show the distributions in the 3D model solar atmosphere. The departure coefficients calculated in the \mtd{} and 1D models are overplotted.}
    \label{fig:departures}
\end{figure*}

To interpret the non-LTE effects, it is useful to look at how they affect the level populations. We illustrate these departures from LTE using the departure coefficients ($b \equiv n_{\mathrm{NLTE}}/n_{\mathrm{LTE}}$), which are shown in \fig{fig:departures}. The figure includes the first six \ion{Ag}{I} levels in order of increasing energy (fine-structure components being in relative LTE in our model and therefore having identical departure coefficients, are only plotted once), as well as the highly excited $5s 5p \, ^4\mathrm{F}_{5/2}$ level and the ground state of \ion{Ag}{II}.

The two diagnostic resonance lines are from transitions between the ground level $5s \, ^2\mathrm{S}_{1/2}$ and the first and third excited levels $5p \, ^2\mathrm{P}_{1/2, \,3/2}$. When we look at level populations in LTE, we can see that the majority species is \ion{Ag}{II}, and almost all of the \ion{Ag}{I} atoms lie in the ground state. The departure coefficients reflect that in non-LTE, the population of the ground level is depleted by the two resonance lines via excitations, as we can tell from the dip around log$\tau_R \approx -1$, corresponding to the region where the lines form (see \fig{fig:atmos}), while they overpopulate the first and third excited levels $4d^{10} 5p \, ^2\mathrm{P}_{1/2,\,3/2}$. 
The second and fourth excited levels $5s^2 \, ^2\mathrm{D}_{5/2,\, 3/2}$ show the same pattern for the departure coefficients as the $5p \, ^2\mathrm{P}_{1/2,\,3/2}$ throughout the atmosphere, indicating that efficient collisions push these levels towards relative LTE. The more highly excited states in the doublet system are similarly connected via effective collisions and therefore have very similar departure coefficients (compare the departure coefficients of $6p \, ^2\mathrm{P}_{1/2}$ with the  $5d \, ^2\mathrm{D}_{3/2}$). This network of effective collisions propagates the overexcitation from the resonance lines all the way to the \ion{Ag}{II} levels, causing an overionisation.

The \ion{Ag}{I} $328\,\nm$ and $338\,\nm$ lines thus drive the non-LTE effects in the system.  To further demonstrate this, we carried out a test where the oscillator strengths of both lines were set to zero when calculating the statistical equilibrium.  Doing so, the statistical equilibrium shifts much closer to LTE. When subsequently calculating the emergent spectra based on these converged populations, the \ion{Ag}{I} $328\,\nm$ and $338\,\nm$ lines appear much deeper, with the difference between LTE and non-LTE equivalent widths being only $0.02\,\dex$ in the \mtd{} model atmosphere.

The departure coefficients of the excited level of $6s \, ^2\mathrm{S}_{1/2}$ look slightly different from the other excited levels, exhibiting a flatter trend closer to unity. This trend can be explained by a competition between collisional coupling of this level to more excited levels, versus photon losses in the relatively strong \ion{Ag}{I} $768.8\,\nm$ and $827.5\,\nm$ doublet, which connect this level to the $ 5p \, ^2\mathrm{P}_{1/2,\,3/2}$ levels (with $\log gf=-0.48$ and $-0.15$ in the present model, respectively). Switching this doublet off, the departure coefficients of the $6s \, ^2\mathrm{S}_{1/2}$ would appear qualitatively similar to those of the more excited levels such as the  $6p \, ^2\mathrm{P}_{1/2}$ and  $5d \, ^2\mathrm{D}_{3/2}$.

Finally, we also show the departure coefficients of the $4d^9 5s 5p \, ^4\mathrm{F}_{5/2} $ level, to illustrate the case of the $4d^9 5s 5p$ levels where the $4d^{10}$ core is excited. As explained in Section~\ref{model_atom}, in the model atom the $4d^9$ levels in the quartet system only couple to the levels in the doublet system via collisions.  There exists efficient coupling via electron collisions between the $4d^9$ quartet levels and the $4d^{10}$ levels of similar excitation energies.  This competes with efficient coupling via hydrogen collisions between the $4d^9$ quartet levels and the $4d^9 5s^2$ levels in the doublet system. The result of this competition is that the departure coefficients of the $4d^9$ quartet levels appear in between those of the excited $4d^{10}$ levels (which appear similar to those for the $6p \, ^2\mathrm{P}_{1/2}$ and  $5d \, ^2\mathrm{D}_{3/2}$ in \fig{fig:departures}), and those of the $4d^9 5s^2$ levels in the doublet system.

The abundances inferred from the different spectrum synthesis models are given in \tab{tab:abundances}. Based on that, the line-averaged abundance correction $\mathrm{\Delta^{I}(1N-1L)}$ is $+0.21\,\dex$. The abundance corrections, here based on measured equivalent widths, are very close for both lines, which can be understood by noting that they connect the same levels (just different fine-structure) and have similar formation depths. Taking non-LTE into account, both diagnostic lines become weaker, which in turn gives a larger non-LTE abundance. The weakening of the lines in non-LTE are the result of the combination of an opacity effect, where the lines opacity decreases due to the non-LTE population of the lower level becoming smaller than the LTE value ($b<1$ in \fig{fig:departures}, first panel); and the source function effect, where the line source function increases due to $b_u/b_l > 1$ where the lines form.

\subsection{3D effects}\label{3d}

\begin{figure*}
\centering
\includegraphics[scale=0.3]{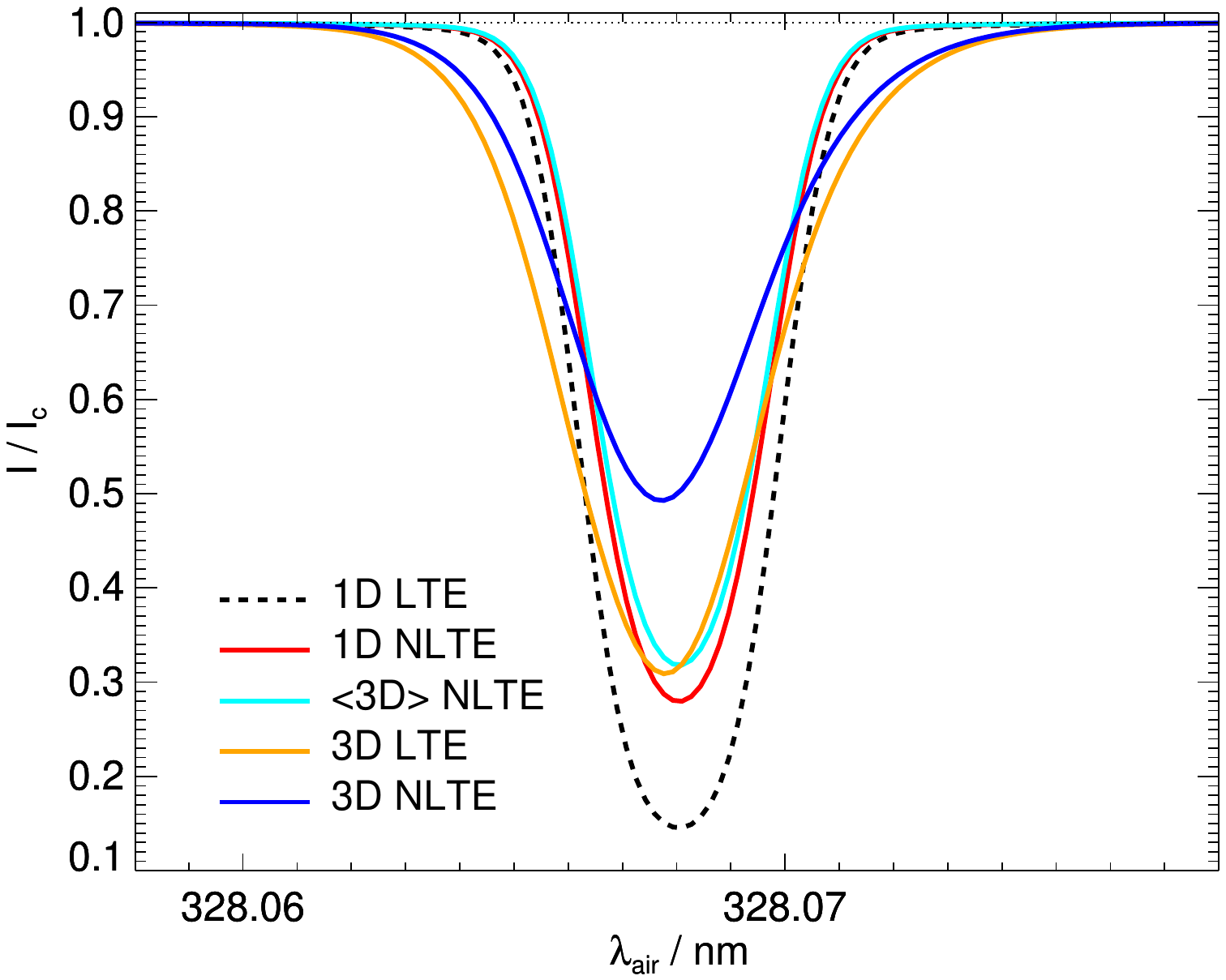}\includegraphics[scale=0.3]{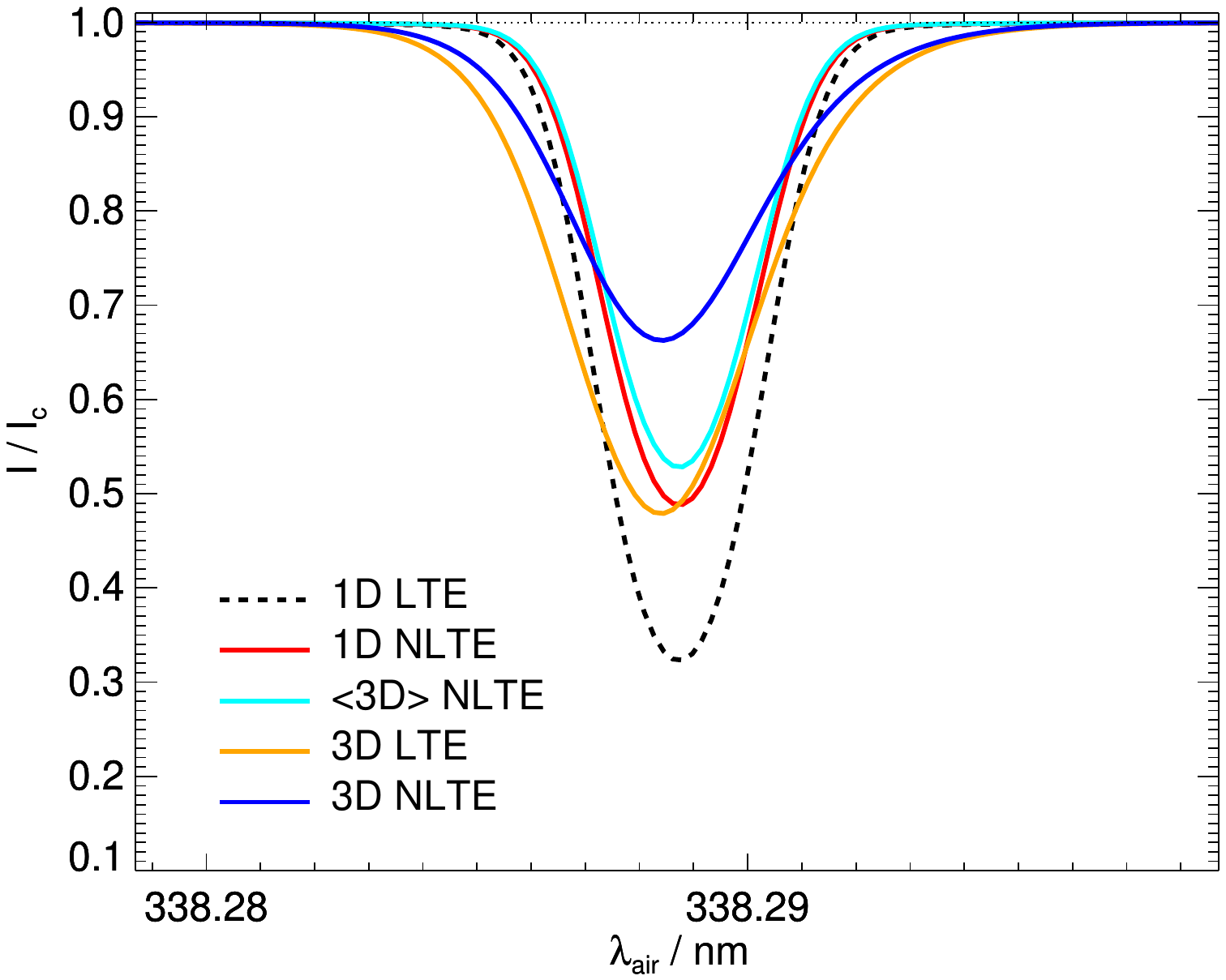}
\caption{Synthetic line profiles of the diagnostic \ion{Ag}{I} lines computed in LTE and non-LTE using different solar model atmospheres. No macroturbulence was added to the lines.
\label{fig:lines}}
\end{figure*}

The total 3D effect is quantified by $\mathrm{\Delta^{I}(3N-1N)}$. To help our understanding, we separate this 3D effect into two components: the effect of the granulation, called the direct effect, and the mean stratification effect, called the indirect effect \citep[e.g.][]{lind_review_2024}.

The direct effect can be quantified by the difference between the 3D and \mtd{}
non-LTE abundances, $\mathrm{\Delta^{I}(3N-\mtdmath{}N)}$, since granulation is
present in the 3D solar model atmosphere but absent in the \mtd{} model
\citep[e.g.][]{2011SoPh..268..255C}. We find that the two diagnostic lines
exhibit different granulation effects, which we attribute to their different
equivalent widths and hence their positions on the curve of growth. In Figure 6b
of \cite{lind_review_2024}, lines that approach the saturated part of the curve
of growth show a rapid increase towards more positive
$\mathrm{\Delta^{I}(3N-\mtdmath{}N)}$. In our case, the $328\,\nm$ line is the
stronger of the two, with a reduced equivalent width of $-5.0$, and yields
$\mathrm{\Delta^{I}(3N-\mtdmath{}N)}=+0.03\,\dex$, consistent with lines of other minority neutral species at similar reduced equivalent width in \cite{lind_review_2024}. By contrast, the weaker $338\,\nm$ line, with a reduced equivalent width 
of $-5.2$, shows a much smaller correction of $+0.01\,\dex$.
The line-average granulation effect is $\mathrm{\Delta^{I}(3N-\mtdmath{}N)} =
+0.02\,\dex$.
This positive granulation effect may reflect that the adopted microturbulence $\vmic$  in the \mtd{} model is too high, overestimating the desaturation effect of the 3D velocity fields.

Meanwhile the indirect effect can be quantified via the line-averaged $\mathrm{\Delta^{I}(\mtdmath{}N - 1N)}$ abundance correction. As the \mtd{} model has a shallower temperature gradient, the \mtd{} line profile is weaker than the 1D, and the abundance difference is always 
positive: $\mathrm{\Delta^{I}(\mtdmath{}N - 1N)} = +0.06\,\dex$, again in agreement with the values for minority species in the Sun in the $\Delta_{\mathrm{strat}}$ vs. $\log_{10} (\mathrm{W_{\lambda}/\lambda)}$ plot from Figure 6a in \cite{lind_review_2024}.  The indirect effect is stronger than the direct effect, thus dominating the overall 3D effect. They both contribute to weakening the 3D non-LTE line, and therefore add up to give a positive line-averaged abundance correction $\mathrm{\Delta^{I}(3N - 1N)}$ 
of $+0.07\,\dex$.

Finally, we illustrate the synthetic line profiles in \fig{fig:lines}. The 3D profiles have broader wings than the  1D and \mtd{}, for a fixed $\vmic$ of $1\,\kms$. One should note that no macroturbulence was added to the 1D and \mtd{} line profiles, which contributes to the differences relative to the 3D line profile, particularly in the wings. Moreover, the 3D line profiles are slightly blueshifted relative to the 1D and \mtd{} line profiles, and clearly display a C-shape asymmetry.  Both of these features are
a consequence of integrating the lines over the surface of the 3D model, sampling both the hot, up-flowing granules and cool, down-flowing lanes \citep[e.g.][]{1982ARA&A..20...61D,2000A&A...359..729A}.

\subsection{Coupling between 3D and non-LTE effects}

We have discussed the non-LTE and 3D effects individually in Sections~\ref{nlte} and~\ref{3d}, but, in reality, the two can interact. In the literature, it is common to see that the 3D LTE and 1D non-LTE abundance corrections are separately applied to the 1D LTE abundance (see, e.g., Table 1 of \citealt{2021A&A...653A.141A}). However, in the case of a coupling between the 3D and non-LTE effects, this procedure can break down, and this approximate ``3D $+$ non-LTE'' abundance may differ by up to $0.5\,\dex$ from the fully consistent 3D non-LTE result, at least in metal-poor stars \citep{2023A&A...672A..90L}.

To quantify this 3D/non-LTE coupling, we compare the line-averaged $\mathrm{\Delta^{I}(3N - 3L)}$ and $\mathrm{\Delta^{I}(1N - 1L)}$ abundance corrections. We find $\mathrm{\Delta^{I}(3N - 3L)} - \mathrm{\Delta^{I}(1N - 1L)} = 0.07\,\dex$, indicating that there is a coupling between the 3D and non-LTE effects for \ion{Ag}{I} in the Sun, with the 3D models yielding larger non-LTE corrections. 

A plausible explanation is that the steeper temperature gradients within the
inhomogeneous 3D radiation-hydrodynamical model atmosphere enhance the non-local
radiation field, leading to more photon pumping in the UV resonance lines,
analogous to that described for \ion{Fe}{I} in 3D models of metal-poor stars
\citep[e.g.][]{2016MNRAS.463.1518A}. This scenario is further supported by the
positive $\mathrm{\Delta^{I}(3N - 3L)} - \mathrm{\Delta^{I}(\mtdmath{}N -
\mtdmath{}L})$ 
difference ($+0.07\,\dex$), showing the enhancement of the non-LTE
departures when moving to the full, inhomogeneous 3D model.

Overall, the coupled 3D and non-LTE effects lead to a severe line-averaged abundance correction 
of $\mathrm{\Delta^{I}(3N - 1L)} = +0.28\,\dex$.  We briefly note that our calculated abundance correction is similarly large in the disc-integrated flux ($\mathrm{\Delta^{F}(3N - 1L)} = +0.29\,\dex$). Both the 3D and non-LTE departures act in the same direction on the lines' equivalent widths, producing positive abundance corrections, but the non-LTE effect is more severe than the 3D effect.  We thus find that 1D LTE is the poorest approximation, while 1D non-LTE is closer to the full 3D non-LTE result than 3D LTE.

\section{Discussion}
\label{discussion}

\subsection{Sensitivity to the radiative and collisional data}
\label{discussion_sensitivity}

\begin{figure*}
\centering
\includegraphics[scale=0.3]{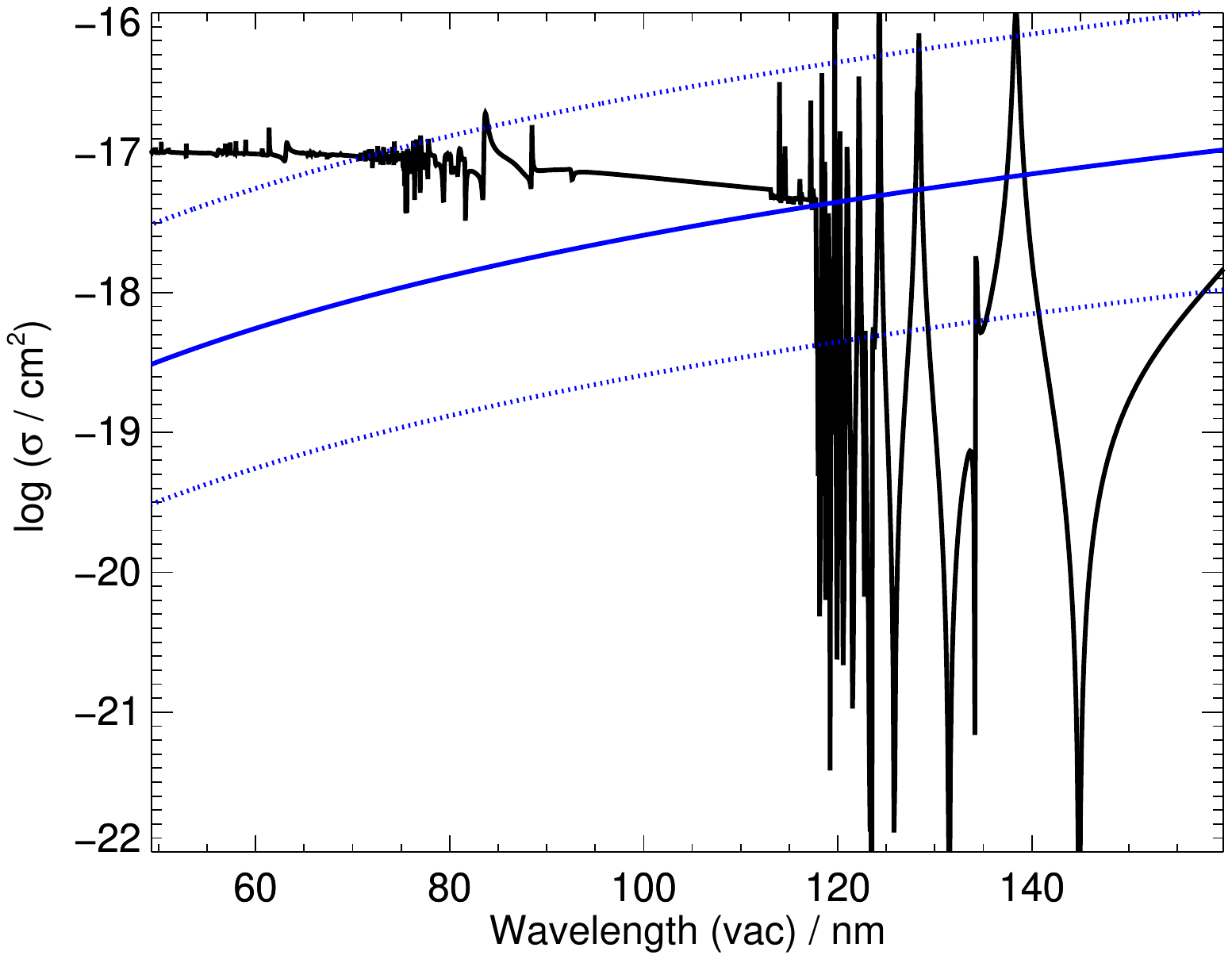}\includegraphics[scale=0.3]{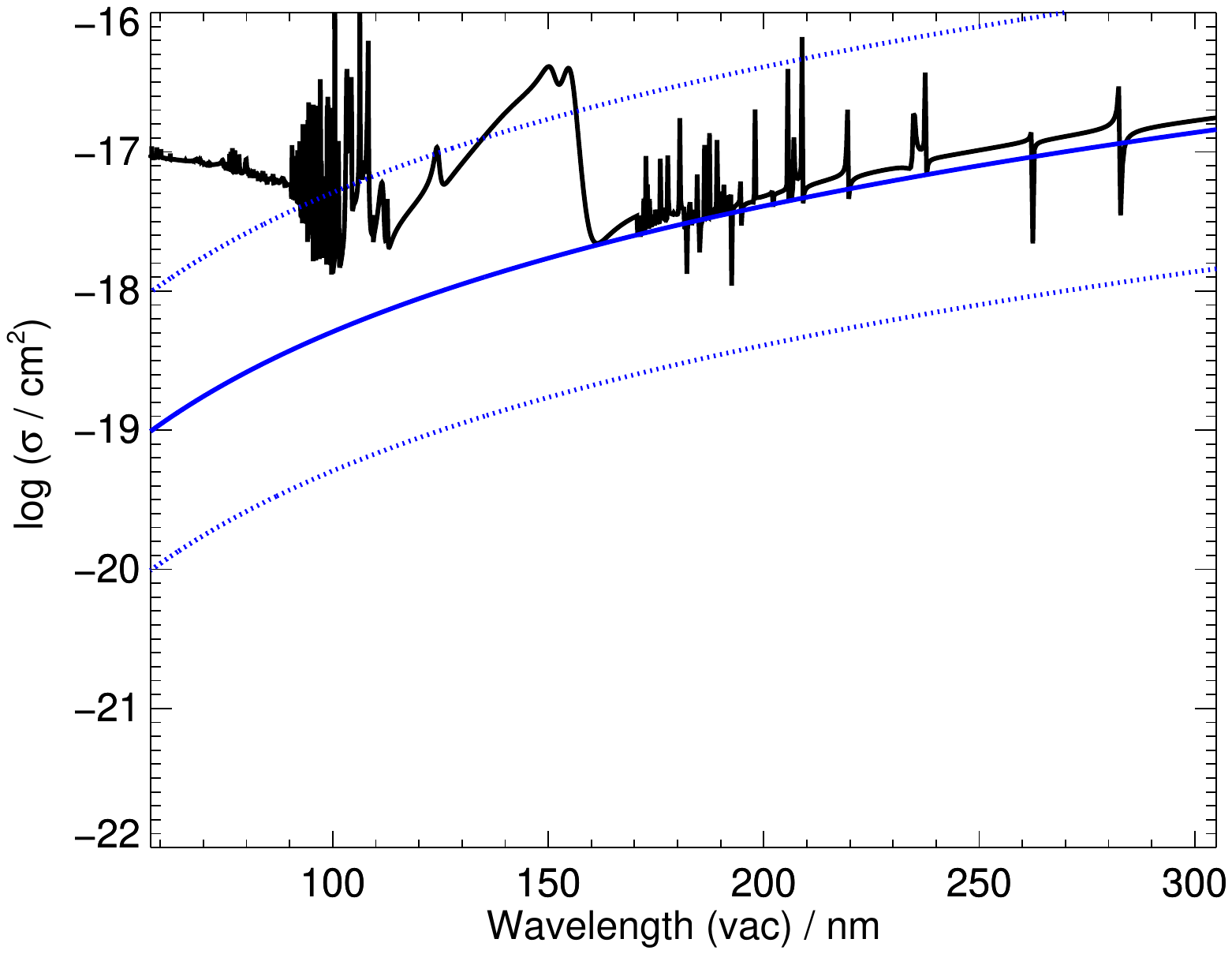}
\caption{Photoionisation cross-sections for neutral copper
from the ground ($\mathrm{3d^{10}4s}$; left panel) and second excited ($\mathrm{3d^{10}4p}$; right panel) states. The black curves show the R-matrix cross-sections 
from \citet{liu_copper_2014}, while the blue curve shows the hydrogenic approximation; the blue dashed curves indicate the hydrogenic cross-sections shifted by $\pm1\,\dex$.
\label{fig:photoionisation}}
\end{figure*}

In order to estimate the sensitivity of non-LTE abundances to uncertainties in the model atom, we performed several tests by varying the atomic data in the model atom. The results of our tests are summarised in \fig{fig:test}, where we show the difference of the equivalent widths in the fiducial model vs. the test model for the two \ion{Ag}{I} diagnostic lines in the \mtd{} solar model atmosphere for a fixed silver abundance of $0.94\,\dex$. These differences are shown for both lines, although they show similar sensitivities as expected, as they belong to the same fine structure multiplet and are of similar strength.

\begin{figure}
\centering
\includegraphics[width=\hsize]{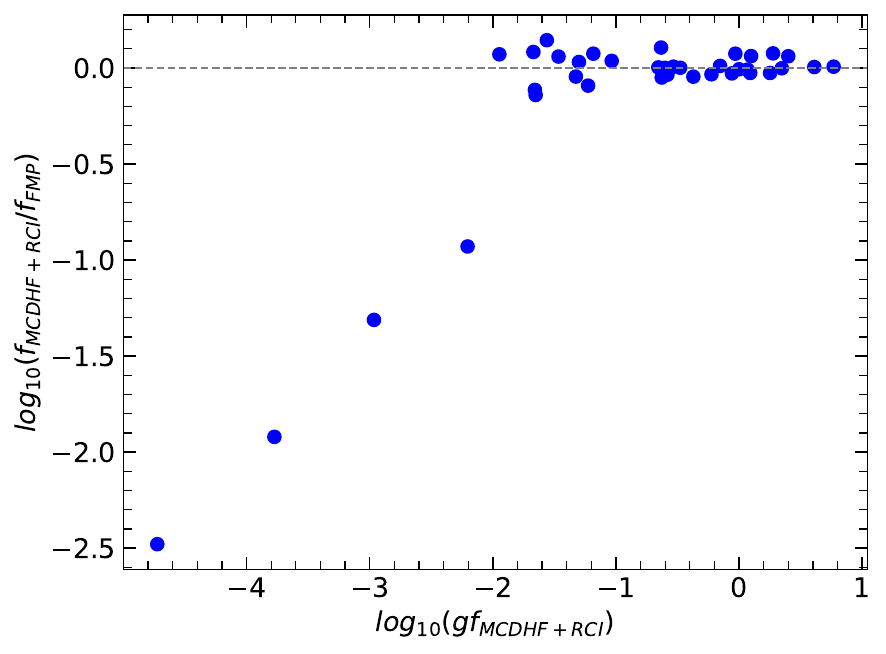}
\caption{Comparison of the newly calculated MCDHF+RCI oscillator strengths with oscillator strengths from \cite{civis_time-resolved_2010} based on the FMP approach.
\label{fig:loggf}}
\end{figure}

\begin{figure}
\centering
\includegraphics[width=\hsize]{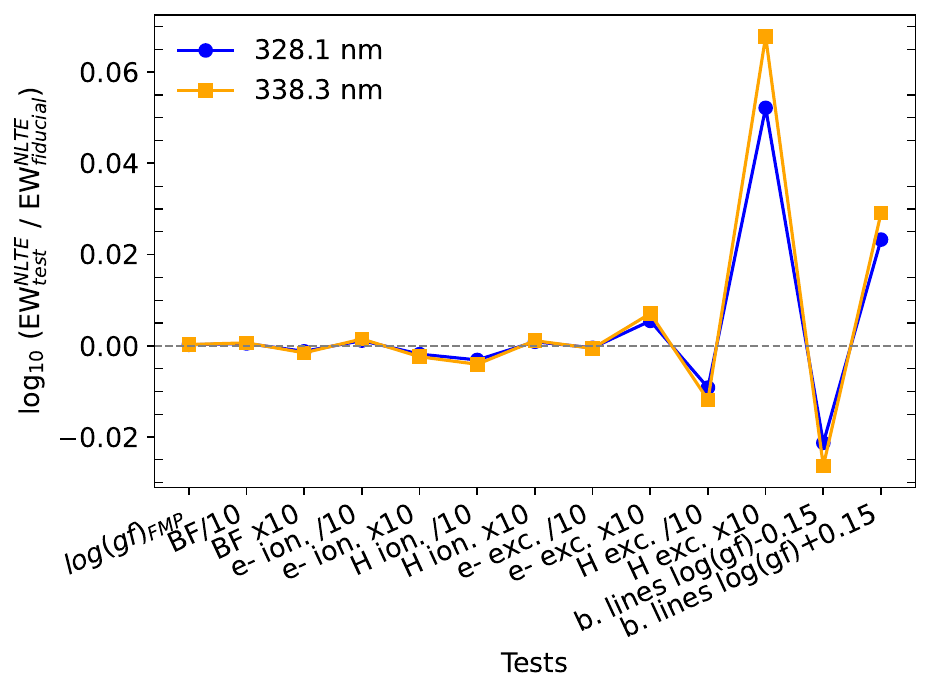}
\caption{Changes in the \mtd{} non-LTE equivalent widths of the two \ion{Ag}{I} diagnostic lines when varying atomic data in the model atom and the background lines (``test'') compared to the fiduciary model.
\label{fig:test}}
\end{figure}

First, we tested the impact of the newly calculated MCDHF+RCI oscillator strengths by replacing them with the less certain oscillator strengths from \cite{civis_time-resolved_2010} when possible, corresponding to ``$\log gf_{\mathrm{FMP}}$'' in \fig{fig:test}. We ended up switching the $f$ values of 40 bound-bound transitions out of 183, with the seven experimental oscillator strengths from NIST remaining unchanged. This had little to no effect on the non-LTE equivalent widths of the \ion{Ag}{I} $328\,\nm$ and $338\,\nm$ lines. Indeed, when we compare the MCDHF+RCI and FMP oscillator strengths in \fig{fig:loggf}, we see a quite good agreement for the strongest lines that can have the largest impact on the statistical equilibrium, as also shown in \cite{2026A&A...709A..31J}. Therefore switching between these oscillator strengths does not significantly change the abundances inferred from the \ion{Ag}{I} $328\,\nm$ and $338\,\nm$ lines.

To estimate the impact of the photoionisation cross-sections on the non-LTE solution, it would be useful to compare the hydrogenic approximation with cross-sections obtained using a more sophisticated method. Such a comparison would also provide good grounds to discuss how realistic the widely used hydrogenic approximation is. However, since photoionisation data for \ion{Ag}{I} are lacking in the literature, we instead first consider the case of the homologous element \ion{Cu}{I}. In \fig{fig:photoionisation}, we compare hydrogenic cross-sections with the R-matrix calculations of \cite{liu_copper_2014} for the ground and second excited levels of \ion{Cu}{I}. The hydrogenic cross-sections are of the same order as the average R-matrix cross-sections (within a factor of $10$), but they typically fail to capture the detailed structure, such as the resonances.

Motivated by this comparison, we tested the impact of the approximate hydrogenic cross-sections in \ion{Ag}{I} by decreasing and increasing all the cross-sections by a factor of $10$. The resulting non-LTE equivalent widths for the \ion{Ag}{I} $328\,\nm$ and $338\,\nm$ vary only marginally, by $0.005$ (for $10$ times smaller) to $0.001\,\dex$ (for $10$ times larger), for both lines. We therefore conclude that these lines are insensitive to large perturbations of the photoionisation data, and that a more detailed treatment of these processes seems not necessary for the present study. This weak sensitivity can be explained by the close collisional coupling of the \ion{Ag}{I} excited levels to the \ion{Ag}{II} levels, combined with the photon losses caused by the resonance lines, driving the non-LTE effects (as explained in Section~\ref{nlte}), effectively compensating the population changes caused by photoionisation.

Next, we tested the impact of collisional rates by decreasing and increasing the different rates by a factor of $10$. \fig{fig:test} shows that the collisional charge transfer (i.e. both electron and hydrogen ionisation) has a negligible effect on the equivalent widths compared to collisional excitation, with the biggest impact from the hydrogen ionisation with a $\Delta (\mathrm{EW})$ of $-0.003\,\dex$. A factor of ten increase in the electron-impact collisional excitations also has a negligible impact on the equivalent widths of the diagnostic lines ($+0.005\,\dex$).

Finally, as explained in Section~\ref{model_atom}, new Ag + H collisional rates have been calculated for the first time using the asymptotic and free electron models. We found that a factor of $10$ increase in the collisional rates has the largest impact on the non-LTE equivalent widths, making the model more sensitive to the hydrogen collision data than to any of the other aforementioned ingredients. \cite{barklem_review_2016} argues that the asymptotic methods are accurate to within a factor of $10$. 
If this factor of $10$ is a realistic uncertainty for the asymptotic hydrogen collisional cross-sections, then we can quantify the corresponding abundance uncertainty.  
The line-averaged abundance correction $\mathrm{\Delta^{I}(\mtdmath{}N -
\mtdmath{}L)}$ went from $0.20\,\dex$ down 
to $0.12\,\dex$ when the hydrogen
collisions were increased, and up to $0.22\,\dex$ when they were decreased.
Taking half the range, we estimate a $0.05\,\dex$ contribution to the overall abundance uncertainty.

This difference in the equivalent widths is largely driven by two collisional transitions: $4d^{10} 5s \, ^2\mathrm{S}_{1/2} \xrightarrow{} 4d^{10} 5p \, ^2\mathrm{P}_{1/2, \,3/2}$, which makes sense given that these are also the transitions behind the diagnostic lines and compete directly against the overexcitation effect discussed in \sect{nlte}. The rates for these transitions are zero in the asymptotic model and relatively strong in the free electron model, highlighting the importance of the latter to avoid missing important collisional transitions
\citep[e.g.][]{2018A&A...616A..89A}.
We further discuss the reliability on the hydrogen collision data in \sect{sun}.

\subsection{Sensitivity to the background line opacities}\label{background_lines}

The background line opacities are not calculated internally, but are pre-calculated in LTE assuming a solar chemical composition. We cannot ignore the impact of the background lines, especially of those that overlap with the \ion{Ag}{I} diagnostic lines, as these drive the non-LTE effects, and are heavily affected by background lines.

It is important to assess the sensitivity of the non-LTE results to the modelling of these background lines. To do so, we varied the $\log gf$'s of all background lines by $\pm 0.15\,\dex$, which roughly corresponds to an uncertainty of ``D'' rank in the NIST accuracy classification scheme. In \fig{fig:test}, we show the effect on the non-LTE equivalent widths. 

Including more background line opacity (by increasing the $\log gf$'s of background lines by $0.15\,\dex$) reduces the mean radiation field in the UV and thus the rate of photoexcitation through the \ion{Ag}{I} $328\,\nm$ and $338\,\nm$ that drives the overexcitation, further reducing the non-LTE effect by increasing their equivalent widths (driving them closer to LTE).

We also quantified the corresponding uncertainty on the abundance correction:
the line-averaged abundance correction $\mathrm{\Delta^{I}(\mtdmath{}N -
\mtdmath{}L)}$ went from $0.20\,\dex$ down to $0.17\,\dex$ when $\log gf$
is increased by $0.15\,\dex$, and up to $0.23\,\dex$ when $\log gf$ is reduced
by $0.15\,\dex$.
This shift of $\pm0.03\,\dex$ reflects an upper limit on the uncertainty due to
the treatment of background line opacities.  While non-negligible, this contribution to the overall abundance uncertainty is smaller
than that due to the inelastic hydrogen collisions that is
estimated to be around $0.05\,\dex$ (\sect{discussion_sensitivity}).

\subsection{The abundance of silver in the Sun}\label{sun}

\begin{figure*}
\centering
\includegraphics[scale=0.21]{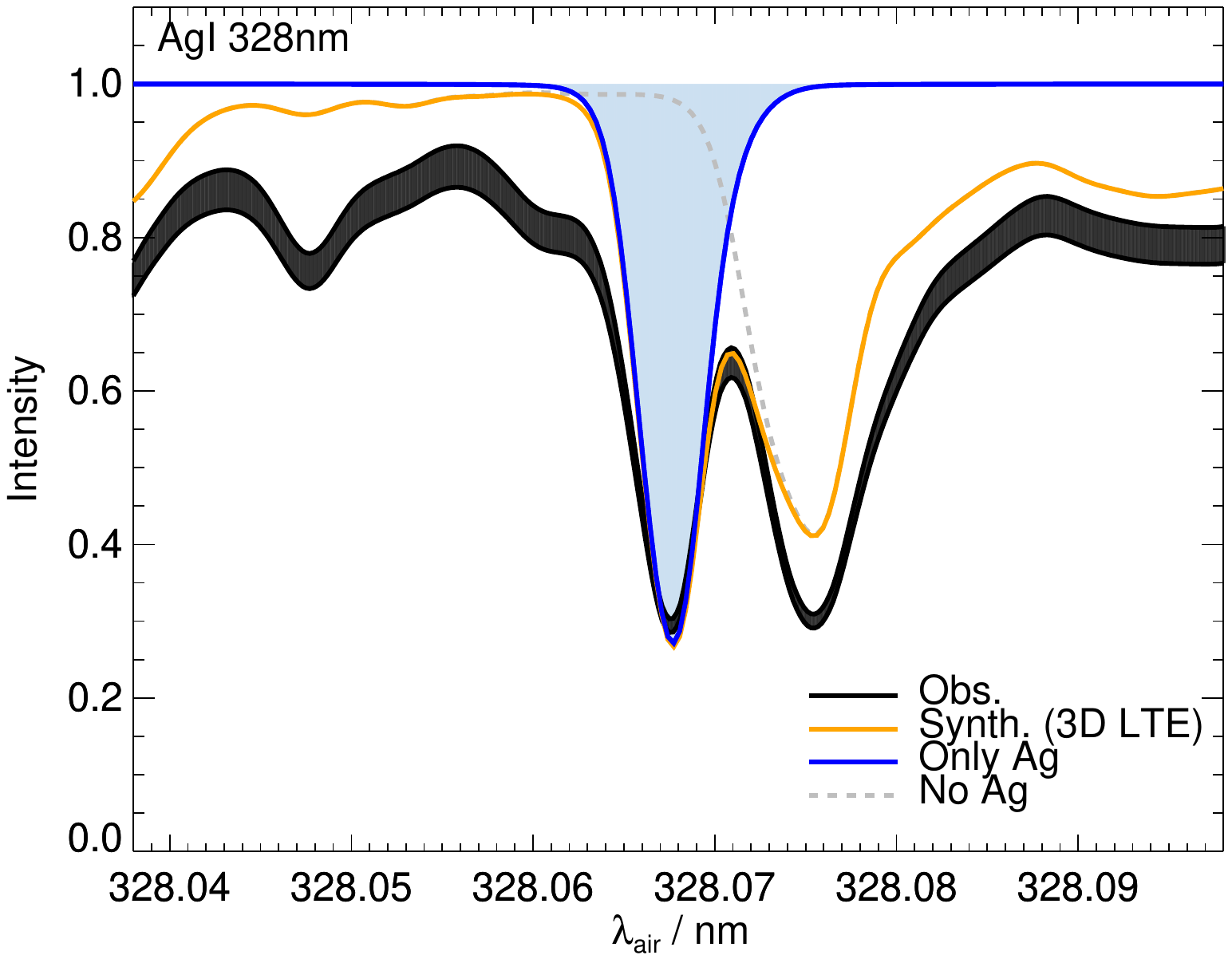}
\includegraphics[scale=0.21]{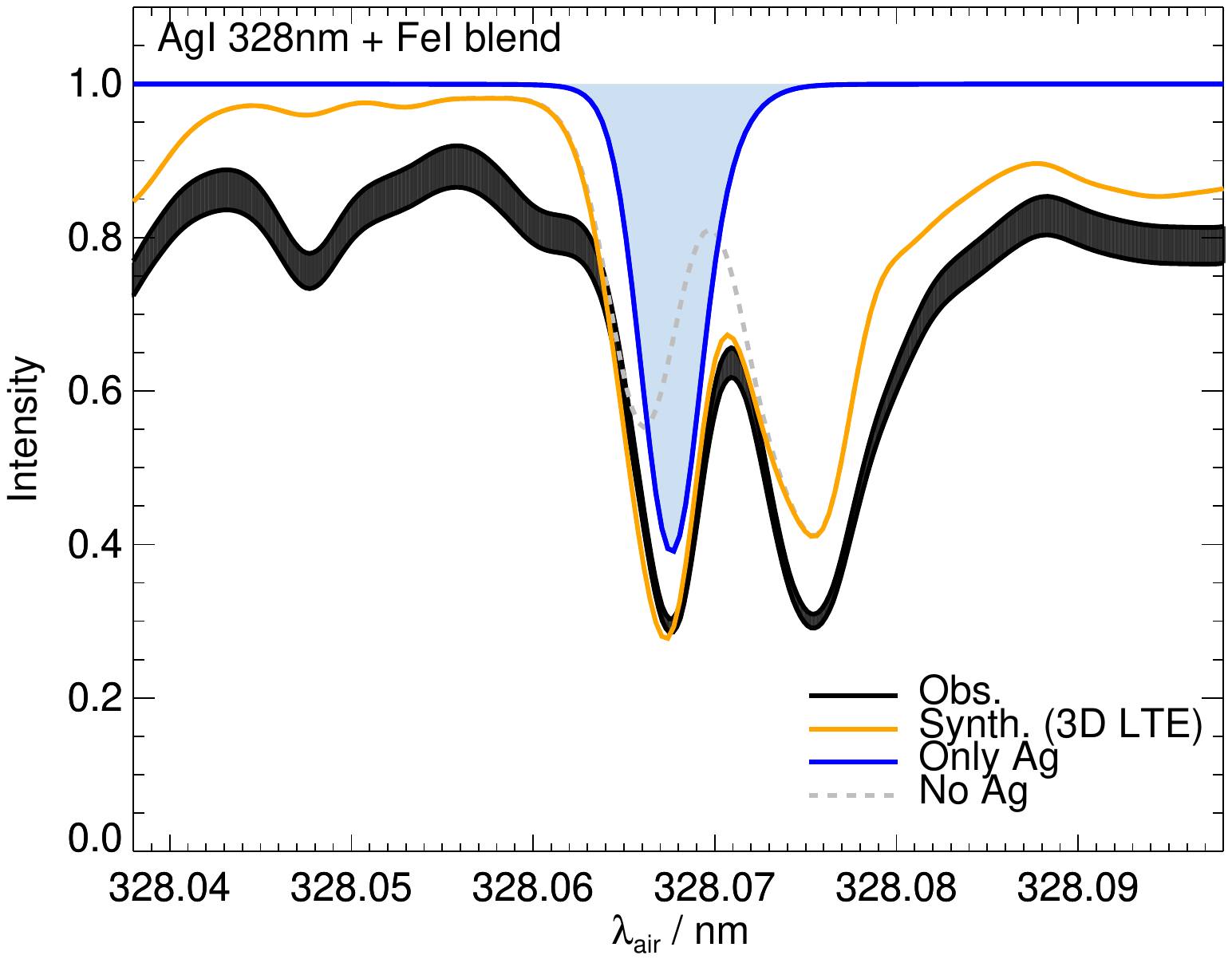}
\includegraphics[scale=0.21]{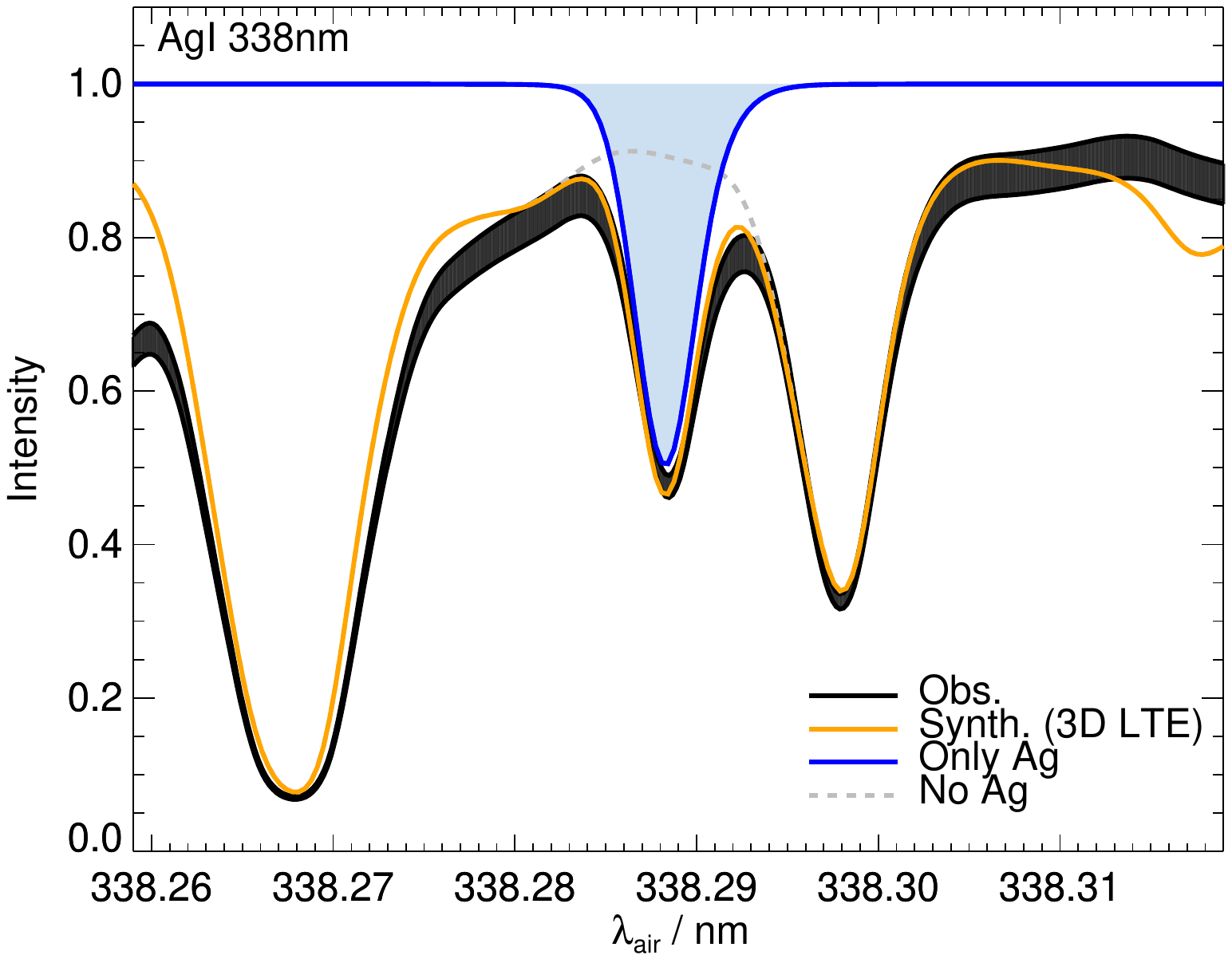}
    \caption{Fits to the Li\`{e}ge disc-centre intensity spectrum,
    with 3D LTE spectrum synthesis and the VALD linelist.
    Shaded black area reflects the uncertainty in placing the continuum.
    Shaded blue area shows the equivalent width of the pure
    \ion{Ag}{I} line.
    Left and middle panels
    show the synthesis of the
    \ion{Ag}{I} $328\,\nm$ line
    omitting and including a predicted 
    \ion{Fe}{I} blend,
    respectively; the recommended equivalent width
    for this line in \tab{tab:abundances} is 
    based on the mean of these two results.}
\label{fig:spectrumfit}
\end{figure*}

\begin{figure*}
\centering
\includegraphics[scale=0.3]{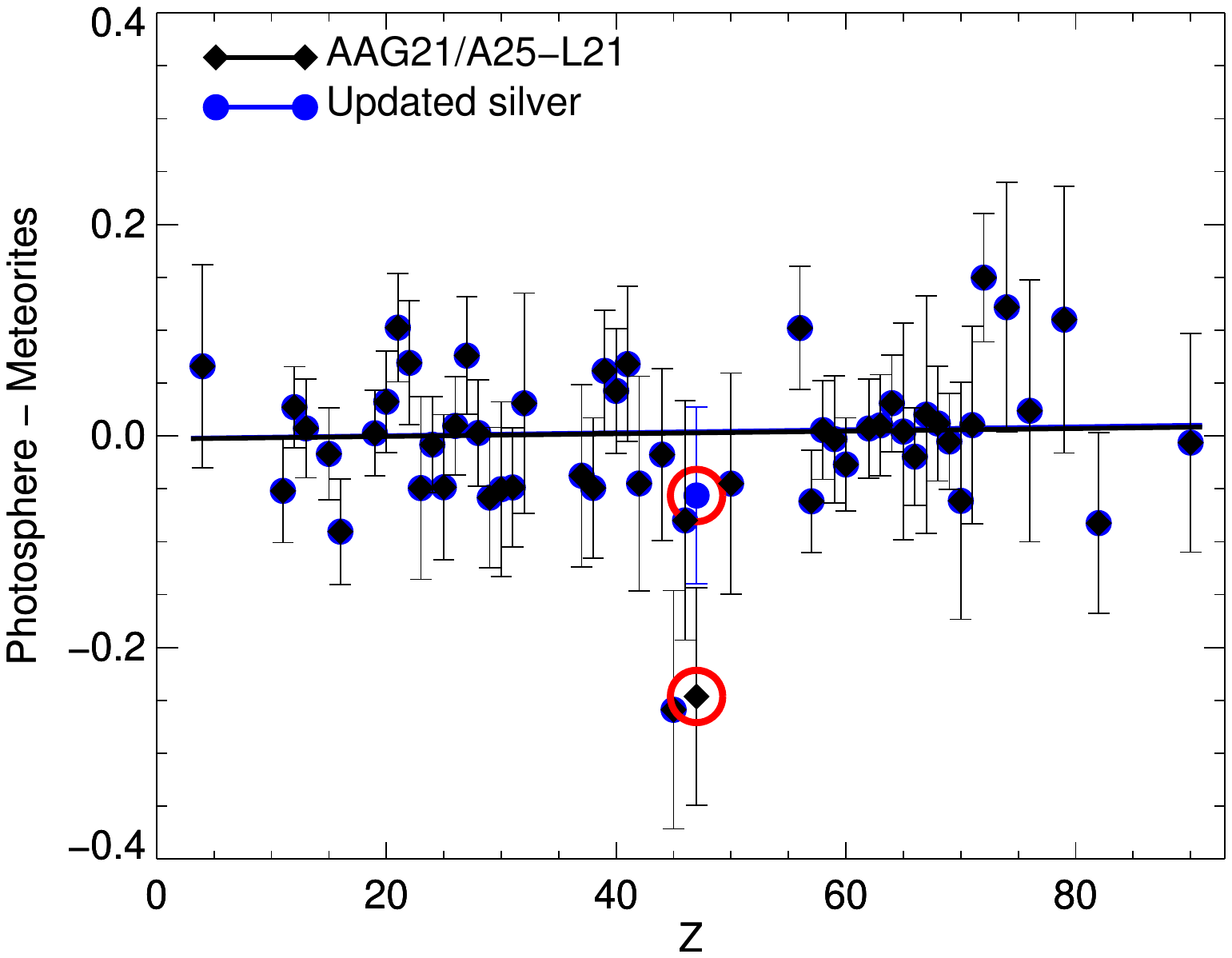}
\includegraphics[scale=0.3]{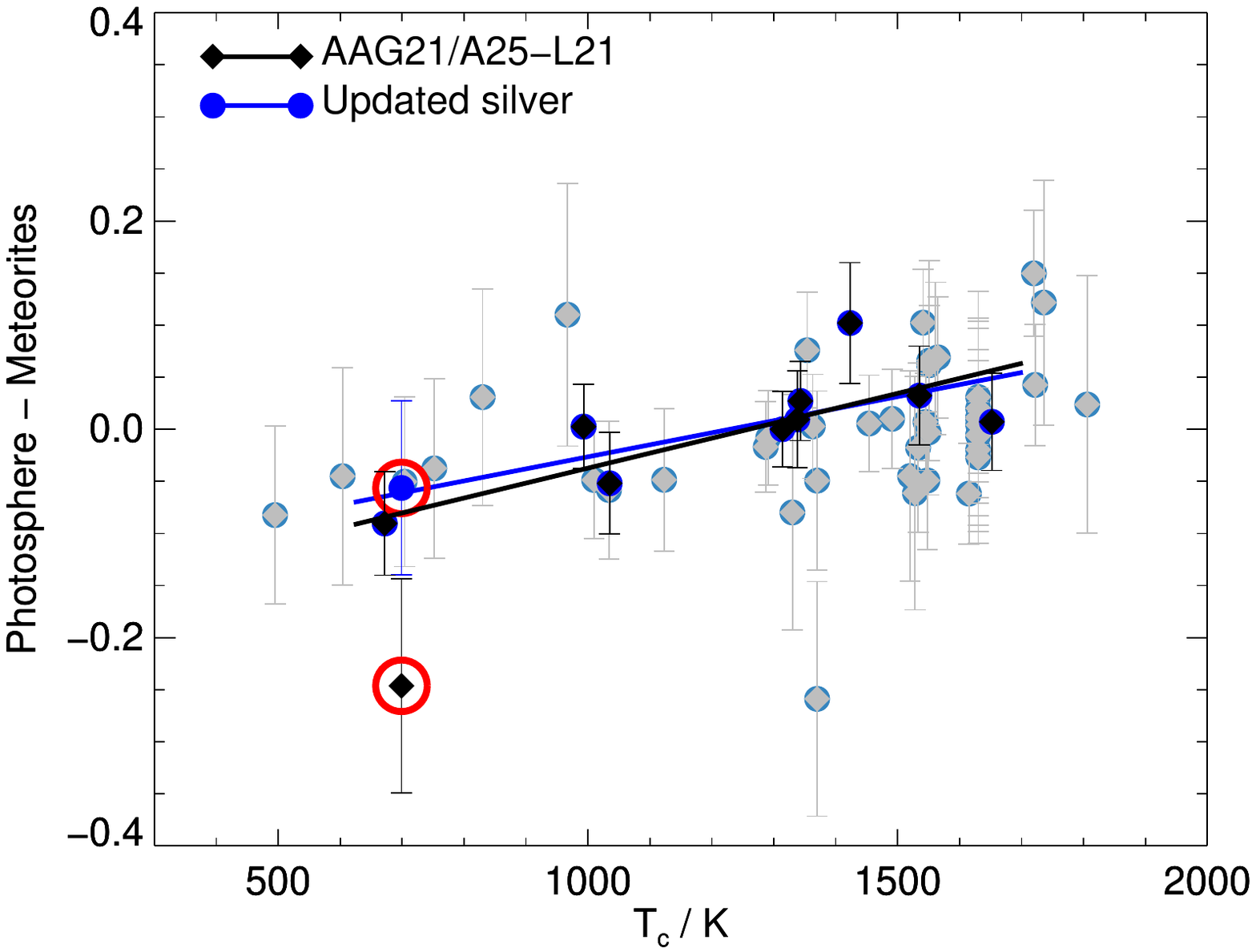}
\caption{Photospheric versus CI chondrite abundance differences as a function of atomic number Z (left panel) and 50$\%$ condensation temperature $T_c$ (right panel), with photospheric abundances from \cite{2021A&A...653A.141A}
and \citet{amarsi_sulphur_2025} (``AAG21/A25''). Silver is circled in red, and the recommended solar silver abundance was used to update AAG21. Weighted linear regressions are overplotted.}
\label{fig:sunmeteorites}
\end{figure*}

The solar silver abundance was measured by \cite{grevesse_elemental_2015} in 3D
LTE, based on equivalent width measurements.  The authors found
$\lgeps{Ag}=1.044$ and $0.869$ from the \ion{Ag}{I} $328\,\nm$ and $338\,\nm$
lines, respectively, leading to a recommended value of $\lgeps{Ag}=0.96\pm0.10$.  
On first look, it is plausible that 
 the relatively large abundance difference between the two
lines, of $0.175\,\dex$, and the large difference of $0.25\,\dex$ with the
meteoritic value ($1.21\,\dex$, via \citealt{lodders_meteoritic_2021} and assuming $\lgeps{Si}=7.51$ from \citealt{2021A&A...653A.141A})
could be explained by coupled
3D non-LTE effects.  Here, we wish to explore how 3D non-LTE effects do indeed
alter the abundance analysis, and provide an updated recommended value for the
silver abundance in the Sun; we discuss the comparison with meteorites in
\sect{sunvsCI}.

We first attempted to verify literature measurements of the
equivalent widths for the 
\ion{Ag}{I} $328\,\nm$ and $338\,\nm$ lines. To do so, 
we performed a 3D LTE synthesis
including nearby blends
using \scate{} \citep{2011A&A...529A.158H}.
The abundances of elements other than silver
were fixed to those of \citet{2021A&A...653A.141A},
and, following \citet{2024A&A...690A.128A}, we included
a correction factor of $1.15$ to the continuous opacity.
The linelist was the default one taken from VALD \citep{1995A&AS..112..525P},
albeit with the \ion{Ag}{I} lines updated to the $\log gf$ values of 
\cite{5944TP}, with hyperfine structure and isotopic contributions
following \citet{hansen_silver_2012},
as well as some small modifications to the blends
as discussed below.
The synthetic spectrum was convolved with a Gaussian kernel
corresponding to resolving power $R=10^{5}$,
and then compared against the Li\`{e}ge disc-centre intensity
atlas \citep{1973apds.book.....D} as illustrated
in \fig{fig:spectrumfit}.

The \ion{Ag}{I} $338\,\nm$ line is less saturated and less blended than the $328\,\nm$ line, and is therefore a more
reliable diagnostic. For the synthesis of this line,
two minor modifications were made to the linelist.
First, the wavelength of the \ion{Cr}{II} line in the blue wing was shifted slightly, from $338.2675\,\nm$ to $338.2679\,\nm$ (in air), to improve the fit to the observed spectrum.
Second, the \ion{Fe}{I} $338.2986\,\nm$ line \citep{1966sst..book.....M} was added to the linelist, adopting the same excitation potential ($2.6085\,\mathrm{eV}$) and a similar oscillator strength ($-2.75\,\dex$) as for the nearby \ion{Fe}{I} $338.3365\,\nm$ line, since this reproduced the observations fairly well. 
We then varied the silver abundance to reproduce the observed spectrum and, from this, estimated the equivalent width of the unblended \ion{Ag}{I} $338\,\nm$ line to be $2.18\,\mathrm{pm}$, with an uncertainty of $0.11\,\mathrm{pm}$ due to the continuum placement. This value is slightly lower than the directly measured values of $2.23\,\mathrm{pm}$ reported by \citet{grevesse_elemental_2015}, and $2.2\,\mathrm{pm}$ by \citet{1966sst..book.....M}. We also carried out a new direct measurement, obtaining $2.35\,\mathrm{pm}$. The difference between that value and our new estimated equivalent width, $2.35-2.18=0.17$, reflects the combined uncertainty associated with the treatment of blends and continuum placement, and we adopt this as the uncertainty in our final equivalent width: $2.18\pm 0.17\,\mathrm{pm}$.

The \ion{Ag}{I} $328\,\nm$ line is more saturated and more 
heavily blended than the $338\,\nm$ line, and is therefore a less reliable abundance diagnostic. As such, we did not attempt to fine-tune the VALD linelist, and instead assigned a larger uncertainty to its equivalent width. In particular, the VALD database includes a predicted \ion{Fe}{I} blend from the Kurucz database \citep{2017CaJPh..95..825K} at $328.0666\,\nm$, with $\log gf=-2.35$ and excitation potential of 3.2830 eV, which strongly overlaps with the \ion{Ag}{I} $328\,\nm$ feature, as shown in \fig{fig:spectrumfit}.
When this predicted \ion{Fe}{I} blend is omitted, the pure equivalent width of the \ion{Ag}{I} $328\,\nm$ line is $3.42\,\mathrm{pm}$, with an uncertainty  of $0.08\,\mathrm{pm}$ due to the continuum placement. This is slightly smaller than the directly measured value of $3.50\,\mathrm{pm}$ give by \citet{grevesse_elemental_2015}. Our own new direct measurement yields a somewhat larger value, $3.74\,\mathrm{pm}$, though still below the $4.4\,\mathrm{pm}$ reported by \citet{1966sst..book.....M}. These differences highlight the difficulty of measuring the equivalent width of this blended feature precisely.
When the predicted \ion{Fe}{I} blend is included, the inferred equivalent width decreases substantially, to $2.69\,\mathrm{pm}$. We therefore adopt the mean of $3.42\,\mathrm{pm}$ and $2.69\,\mathrm{pm}$ and take half their difference as the uncertainty, arriving at $3.06\pm 0.36\,\mathrm{pm}$. This is $0.06\,\dex$ smaller than the value adopted by \citet{grevesse_elemental_2015}; because the line is saturated, the resulting downward revision to the 3D LTE silver abundance is even larger, amounting to $0.15\,\dex$.

Based on the revised equivalent widths, we determined abundances from the \ion{Ag}{I} $328\,\nm$ and $338\,\nm$ lines. Abundances were first derived in 3D LTE using \scate{}, and differential corrections from \balder{} were then applied to infer the abundances for the other models. The resulting values are listed in \tab{tab:abundances}. We recommend the 3D non-LTE value, $1.15\,\dex$, that is the mean weighted by uncertainties of $0.11\,\dex$ and $0.05\,\dex$ on the $328\, \nm$ and $338 \, \nm$ lines respectively, propagated from the equivalent width measurements above.  We combine the weighted observational uncertainty of $0.05\,\dex$ with modelling uncertainties of $0.05\,\dex$ due to inelastic hydrogen collisions (\sect{discussion_sensitivity}) and $0.03\,\dex$ due to the treatment of background line opacities (\sect{background_lines}).  Our recommended result is thus $\lgeps{Ag}=1.15\pm 0.08$.  Although the differences between 3D non-LTE and 3D LTE in \tab{tab:abundances} are $\mathrm{\Delta^{I}(3N-3L)}=0.27\,\dex$, our result is only $0.19\,\dex$ higher than the 3D LTE abundance reported by \citet{grevesse_elemental_2015}; the differences are alleviated by our revised equivalent widths, in particular the inclusion of the predicted \ion{Fe}{I} blend at $328.0666\,\nm$, which significantly lowers the inferred abundance.

In our analysis, the two \ion{Ag}{I} lines give nearly identical abundances, differing by only $0.02,\dex$ in 3D non-LTE. This contrasts with the much larger spread of $0.175,\dex$ obtained by \cite{grevesse_elemental_2015} in 3D LTE. Nevertheless, this close agreement should likely be viewed as fortuitous, given our ad hoc treatment of the predicted \ion{Fe}{I} blend at $328.0666,\nm$. Given that predicted lines from the Kurucz database with $\log gf<-2$ may be in error by around $-0.4\pm0.8\,\dex$ (Section 2.2.2 of \citealt{2017MNRAS.468.4311L}), a stronger constraint on its oscillator strength is required to confirm this result.
We note that it may be worthwhile to
attempt an empirical calibration of this blend
via high-resolution observations of
the centre-to-limb variation \citep[e.g.][]{2024A&A...690A.128A}, for instance
based on forthcoming data from the \texttt{SUNRISE} UV Spectropolarimeter and
Imager (\texttt{SUSI}; \citealt{2025SoPh..300...65F,2025SoPh..300...75K}).
Such an analysis would also help validate
our non-LTE modelling and our treatment
of inelastic collisions with neutral hydrogen
\citep[e.g.][]{2004A&A...423.1109A}.
In all, this would be a promising approach
to reducing the uncertainty on the mean silver abundance below 
$0.08\,\dex$.

\subsection{Comparison with previous results for the Sun and CI chondrites}\label{sunvsCI}

In Figures 5 and 6 of \cite{2021A&A...653A.141A}, the authors show a comparison of the recommended solar photospheric abundances with those from CI chondrites. Out of these elements, silver is a nominal $>2\sigma$ outlier. The authors adopted the 3D LTE solar photospheric silver abundance from \cite{grevesse_elemental_2015} of $\lgeps{Ag}=0.96\pm0.10$. Our goal is to update the solar photospheric abundance they used with the presently determined 3D non-LTE silver abundance and see how it affects these abundance trends.

We updated the plots in \cite{2021A&A...653A.141A} by showing the new
recommended 3D non-LTE silver abundance in \fig{fig:sunmeteorites}. Our plots
show elements for which the combined (photospheric and meteoritic) abundance
uncertainties are less than $0.15\,\dex$.  
We include the updated 3D non-LTE sulphur abundance from \citet{amarsi_sulphur_2025} (referred to as ``AAG21/A25'' in the 
legends).
On the left panel of
\fig{fig:sunmeteorites}, we can see that the new silver abundance reduces the
difference between the photospheric and meteoritic abundance from $-0.25\,\dex$ 
to $-0.06\,\dex$. This reflects the $0.19\,\dex$ difference between our recommended
solar silver abundance and that of \citet{grevesse_elemental_2015} (\sect{sun}).
The photospheric value is now 
consistent with the meteoritic value to within the
estimated uncertainties.

The right panel of \fig{fig:sunmeteorites} shows the difference between the photospheric and meteoritic values as a function of 50$\%$ equilibrium condensation temperature at a pressure of $10^{-4}$ bar for a solar composition gas. For the CI chondrites, the values are from \citet{lodders_meteoritic_2021}, converted to the solar scale using silicon as the reference element.
The 50$\%$ equilibrium condensation temperature were taken from \citet{2019AmMin.104..844W}, in which there are updates of the order several hundred kelvin for the moderately volatile elements (such as silver) compared to the often used data set of \citet{2003ApJ...591.1220L}. On the plot, elements whose solar abundances are derived based on a full 3D non-LTE analysis are highlighted by darker colours. These elements are Na, Mg, Al, Si, K, Ca, Fe and Ba (see Table 1 of \citealt{2021A&A...653A.141A}). 
A linear fit is applied to this 3D non-LTE subset, taking into account the uncertainties (error bars). Changing the reference element would shift all the points up or down, but not affect the fitted gradient.

Whether the differences between the composition of the solar photosphere and of the CI chondrites are a real effect (as suggested by e.g.~\citealt{2010MNRAS.407..314G}, \citealt{2018ApJS..238...11D}, and \citealt{2024M&PS...59.3193J}), or resulting from inaccuracies in the solar spectroscopy \citep[e.g.][]{2025SSRv..221...23L}, is still debated. What we see from the \cite{2021A&A...653A.141A} results 
(referred to as ``AAG21/A25-L21'' in the figure) is that there is a slight trend in the abundance difference as a function of the condensation temperature. Specifically, with silicon as the reference element, 
the moderately volatile elements ($500 \lesssim T_c \lesssim 1250\,\kelvin$) are slightly depleted in the Sun, while the refractory elements ($T_c \gtrsim 1370\,\kelvin$) are enhanced in the Sun; 
the elements in the middle (which includes silicon with $T_c=1314\,\kelvin$)
agree well between the Sun and meteorites.  Silver is a moderately volatile
element, with $T_c = 699\,\kelvin$. When we update the photospheric silver
abundance with our recommended abundance, and including the estimated
uncertainty on the weighted mean abundance,  
silver falls into line with the other
elements that are based on a full 3D non-LTE analysis.
Refining the analysis so as to reduce the uncertainty below
$0.08\,\dex$ (see discussion in \sect{sun}) would be worthwhile to help verify,
or otherwise, the presence of any systematic differences between the solar and
meteoritic abundances.

\section{Conclusions}
\label{conclusion}

We present a model atom for \ion{Ag}{I} and perform a full 3D non-LTE analysis of the solar \ion{Ag}{I} resonance lines for the first time. The model atom is built from physically motivated, carefully curated radiative and collisional data. For bound–bound transitions, we supplement the limited set of experimental oscillator strengths with new calculations. Inelastic hydrogen collisions are computed using a combined asymptotic and free-electron approach to provide a complete dataset.

Because no non-LTE reference calculations for silver exist in the literature, we assessed the robustness of our model through targeted sensitivity tests. The largest sensitivities arise from the hydrogen collisional excitation rates and from the treatment of background opacity in lines overlapping the \ion{Ag}{I} diagnostic resonance features, with the hydrogen collision data remaining the dominant caveat. As an additional check, we compared abundances inferred from disc-centre intensity and disc-integrated flux equivalent widths; their close agreement suggests that the adopted hydrogen collision rates are not in significant error.

We find that departures from LTE are driven by the two diagnostic \ion{Ag}{I}
resonance lines at $328\,\nm$ and $338\,\nm$, leading to overexcitation and
overionisation from the ground level and weakening the lines in non-LTE relative
to LTE. The 3D effect reinforces this, further reducing the line strengths. The
coupling of the 3D and non-LTE effects produces large positive abundance
corrections, 
$\mathrm{\Delta^{I}(3N-1L)}=+0.28\,\dex$ and $\mathrm{\Delta^{F}(3N-1L)}=+0.29\,\dex$. Since the non-LTE effect dominates over the 3D effect, a 1D non-LTE treatment provides a much closer approximation to the full 3D non-LTE result than either 1D LTE or 3D LTE.

Using revised equivalent width measurements of the 328 and 338 nm lines, we re-evaluate the solar silver abundance relative to the 3D LTE value from \cite{grevesse_elemental_2015}. We 
recommend $\lgeps{Ag} = 1.15 \pm 0.08$ in 3D non-LTE,
with measurement and modelling uncertainties
added in quadrature.
This is $0.19\,\dex$ higher than the previous 3D LTE value
($0.96\pm0.10\,\dex$).
The difference is driven by a positive
3D non-LTE versus 3D
LTE abundance correction, that is offset by downwards revisions to 
the equivalent widths of the two \ion{Ag}{I} lines.

The revised solar abundance resolves $0.19\,\dex$
of the $0.25\,\dex$ discrepancy between the
solar photospheric and CI chondrite silver abundances reported by
\cite{2021A&A...653A.141A}, where the photospheric value appeared significantly
underestimated. 
The residual $0.06\,\dex$ discrepancy is consistent
with what has been found for other moderately volatile elements
and may reflect an intrinsic bias in CI chondrites
\citep[e.g.][]{2024M&PS...59.3193J,amarsi_sulphur_2025}.
To confirm this, it would be worthwhile to 
refine the analysis presented here
so as to obtain a more precise measurement
of the solar silver abundance.

We also provide non-LTE departure coefficients\footnote{\url{https://doi.org/10.5281/zenodo.3888393}} for dwarfs and giants across the \marcs{} grid of 1D model atmospheres \citep{2008A&A...486..951G}. Given that the 3D and non-LTE effects act in the same direction (at least for the Sun), a 1D non-LTE approach is expected to be more realistic than 1D LTE for late-type stars. These departure coefficients can therefore be combined with 1D LTE spectrum-synthesis codes such as \texttt{SME} \citep{2017A&A...597A..16P} and \texttt{PySME} \citep{2023A&A...671A.171W} to improve the accuracy of stellar silver abundance analyses.

Finally, we predict that the 3D non-LTE versus 1D LTE abundance corrections for the \ion{Ag}{I} $328 \, \nm$ and $338 \,\nm$ resonance lines become increasingly positive towards lower [Fe/H]. As for neutral iron \citep[e.g.][]{2016MNRAS.463.1518A}, the overexcitation/overionisation mechanism should be strengthened as metal-line opacity decreases, enhancing the UV radiation field. This may be amplified further by the typically steeper temperature gradients in metal-poor 3D models. A likely consequence is that the inferred [Ag/Fe] vs. [Fe/H] may be steeper in 3D non-LTE than in 1D LTE analyses of dwarfs, for example those of \citet{hansen_origin_2011}, particularly if the \ion{Ag}{I} corrections exceed those affecting \ion{Fe}{I}. We will quantify and discuss these effects in a subsequent paper.

\section*{Data availability}
The Ag + H rate coefficients are available
at \url{https://doi.org/10.5281/zenodo.20038646}.  The 1D non-LTE departure coefficients across the \marcs{} model atmosphere grid are available at 
\url{https://doi.org/10.5281/zenodo.20037437}.

\begin{acknowledgements}
We thank the referee for suggestions that helped improve the analysis and manuscript.
We thank Dan Kiselman for providing helpful advice about solar observations.
AMA acknowledges support from the Swedish Research Council (VR 2020-03940, VR 2025-05167) and from the Crafoord Foundation via the Royal Swedish Academy of Sciences (CR 2024-0015). The computations were enabled by resources at the National Supercomputing Centre (NSC, Tetralith cluster) provided by the National Academic Infrastructure for Supercomputing in Sweden (NAISS), partially funded by the Swedish Research Council through grant agreement no. 2022-06725. PJ acknowledges support from the Swedish Research Council (VR 2023-05367). BKS acknowledges ANRF grant no. CRG/2023/002558 and Department of Space, Government of India for financial supports.
\end{acknowledgements}


\bibliographystyle{aa_url} 
\bibliography{bibl.bib}

\end{document}